\documentclass[12pt]{article}
\usepackage{epsfig}
\usepackage{amsmath}
\usepackage{hhline}
\usepackage{amssymb}
\usepackage{times}
\usepackage{cite}
\usepackage{psfrag}
\usepackage{color}
\definecolor{vlMJ}{rgb}{0.2,0.6,0.8}
\definecolor{elMJ}{rgb}{0.53,0.4,0.34}

\newlength{\dinwidth}
\newlength{\dinmargin}
\begin{document}  
\newcommand{\pom}{{I\!\!P}}
\newcommand{\reg}{{I\!\!R}}
\newcommand{\slowpi}{\pi_{\mathit{slow}}}
\newcommand{\fiidiii}{F_2^{D(3)}}
\newcommand{\fiidiiiarg}{\fiidiii\,(\beta,\,Q^2,\,x)}
\newcommand{\n}{1.19\pm 0.06 (stat.) \pm0.07 (syst.)}
\newcommand{\nz}{1.30\pm 0.08 (stat.)^{+0.08}_{-0.14} (syst.)}
\newcommand{\fiidiiiful}{F_2^{D(4)}\,(\beta,\,Q^2,\,x,\,t)}
\newcommand{\fiipom}{\tilde F_2^D}
\newcommand{\ALPHA}{1.10\pm0.03 (stat.) \pm0.04 (syst.)}
\newcommand{\ALPHAZ}{1.15\pm0.04 (stat.)^{+0.04}_{-0.07} (syst.)}
\newcommand{\fiipomarg}{\fiipom\,(\beta,\,Q^2)}
\newcommand{\pomflux}{f_{\pom / p}}
\newcommand{\nxpom}{1.19\pm 0.06 (stat.) \pm0.07 (syst.)}
\newcommand {\gapprox}
   {\raisebox{-0.7ex}{$\stackrel {\textstyle>}{\sim}$}}
\newcommand {\lapprox}
   {\raisebox{-0.7ex}{$\stackrel {\textstyle<}{\sim}$}}
\def\gsim{\,\lower.25ex\hbox{$\scriptstyle\sim$}\kern-1.30ex%
\raise 0.55ex\hbox{$\scriptstyle >$}\,}
\def\lsim{\,\lower.25ex\hbox{$\scriptstyle\sim$}\kern-1.30ex%
\raise 0.55ex\hbox{$\scriptstyle <$}\,}
\newcommand{\pomfluxarg}{f_{\pom / p}\,(x_\pom)}
\newcommand{\dsf}{\mbox{$F_2^{D(3)}$}}
\newcommand{\dsfva}{\mbox{$F_2^{D(3)}(\beta,Q^2,x_{I\!\!P})$}}
\newcommand{\dsfvb}{\mbox{$F_2^{D(3)}(\beta,Q^2,x)$}}
\newcommand{\dsfpom}{$F_2^{I\!\!P}$}
\newcommand{\gap}{\stackrel{>}{\sim}}
\newcommand{\lap}{\stackrel{<}{\sim}}
\newcommand{\fem}{$F_2^{em}$}
\newcommand{\tsnmp}{$\tilde{\sigma}_{NC}(e^{\mp})$}
\newcommand{\tsnm}{$\tilde{\sigma}_{NC}(e^-)$}
\newcommand{\tsnp}{$\tilde{\sigma}_{NC}(e^+)$}
\newcommand{\st}{$\star$}
\newcommand{\sst}{$\star \star$}
\newcommand{\ssst}{$\star \star \star$}
\newcommand{\sssst}{$\star \star \star \star$}
\newcommand{\tw}{\theta_W}
\newcommand{\sw}{\sin{\theta_W}}
\newcommand{\cw}{\cos{\theta_W}}
\newcommand{\sww}{\sin^2{\theta_W}}
\newcommand{\cww}{\cos^2{\theta_W}}
\newcommand{\trm}{m_{\perp}}
\newcommand{\trp}{p_{\perp}}
\newcommand{\trmm}{m_{\perp}^2}
\newcommand{\trpp}{p_{\perp}^2}
\newcommand{\alp}{\alpha_s}

\newcommand{\alps}{\alpha_s}
\newcommand{\sqrts}{$\sqrt{s}$}
\newcommand{\LO}{$O(\alpha_s^0)$}
\newcommand{\Oa}{$O(\alpha_s)$}
\newcommand{\Oaa}{$O(\alpha_s^2)$}
\newcommand{\PT}{p_{\perp}}
\newcommand{\JPSI}{J/\psi}
\newcommand{\sh}{\hat{s}}
\newcommand{\uh}{\hat{u}}
\newcommand{\MP}{m_{J/\psi}}
\newcommand{\PO}{I\!\!P}
\newcommand{\xbj}{x}
\newcommand{\xpom}{x_{\PO}}
\newcommand{\ttbs}{\char'134}
\newcommand{\xpomlo}{3\times10^{-4}}  
\newcommand{\xpomup}{0.05}  
\newcommand{\dgr}{^\circ}
\newcommand{\pbarnt}{\,\mbox{{\rm pb$^{-1}$}}}
\newcommand{\gev}{\,\mbox{GeV}}
\newcommand{\WBoson}{\mbox{$W$}}
\newcommand{\fbarn}{\,\mbox{{\rm fb}}}
\newcommand{\fbarnt}{\,\mbox{{\rm fb$^{-1}$}}}
\newcommand{\dsdx}[1]{$d\sigma\!/\!d #1\,$}
\newcommand{\eV}{\mbox{e\hspace{-0.08em}V}}

\newcommand{\qsq}{\ensuremath{Q^2} }
\newcommand{\gevsq}{\ensuremath{\mathrm{GeV}^2} }
\newcommand{\et}{\ensuremath{E_t^*} }
\newcommand{\rap}{\ensuremath{\eta^*} }
\newcommand{\gp}{\ensuremath{\gamma^*}p }
\newcommand{\dsiget}{\ensuremath{{\rm d}\sigma_{ep}/{\rm d}E_t^*} }
\newcommand{\dsigrap}{\ensuremath{{\rm d}\sigma_{ep}/{\rm d}\eta^*} }

\newcommand{\Rp}{\mbox{$\not \hspace{-0.15cm} R_p$}}
\newcommand{\GeV}{\rm GeV}
\newcommand{\TeV}{\rm TeV}
\newcommand{\pb}{\rm pb}
\newcommand{\cm}{\rm cm}
\newcommand{\hdick}{\noalign{\hrule height1.4pt}}

\def\Journal#1#2#3#4{{#1} {\bf #2} (#3) #4}
\def\NCA{\em Nuovo Cimento}
\def\NIM{\em Nucl. Instrum. Methods}
\def\NIMA{{\em Nucl. Instrum. Methods} {\bf A}}
\def\NPB{{\em Nucl. Phys.}   {\bf B}}
\def\PLB{{\em Phys. Lett.}   {\bf B}}
\def\PRL{\em Phys. Rev. Lett.}
\def\PRD{{\em Phys. Rev.}    {\bf D}}
\def\ZPC{{\em Z. Phys.}      {\bf C}}
\def\EJC{{\em Eur. Phys. J.} {\bf C}}
\def\CPC{\em Comp. Phys. Commun.}

\begin{titlepage}

\noindent
\begin{flushleft}
{\tt DESY 10-181    \hfill    ISSN 0418-9833} \\
{\tt October 2010}                  \\
\end{flushleft}

\vfill
\begin{center}
\begin{Large}

{\bf Search for Squarks in \boldmath{$R$}--parity Violating Supersymmetry 
in \boldmath{$ep$} Collisions at HERA }

\vspace{2cm}

H1 Collaboration

\end{Large}
\end{center}

\vspace{2cm}

\begin{abstract}
A search for squarks in $R$--parity violating supersymmetry is performed
in $e^{\pm} p$ collisions at HERA using the H1 detector.
The full data sample
taken at a centre--of--mass energy $\sqrt{s}=319$~GeV
is used for the analysis, corresponding to an integrated luminosity of 
$255$~pb$^{-1}$ of $e^{+}p$ and $183$~pb$^{-1}$ of $e^{-}p$ collision data.
The resonant production of squarks via a Yukawa coupling $\lambda'$ is considered,
taking into account direct and indirect $R$--parity violating decay modes.
Final states with jets and leptons are investigated.
No evidence for squark production is found and
mass dependent limits on $\lambda'$
are obtained in the framework of the Minimal
Supersymmetric Standard Model and in the Minimal Supergravity Model.
In the considered part of the parameter space, for a Yukawa
coupling of electromagnetic strength $\lambda'= 0.3$,
squarks of all flavours are excluded up to masses of $275$~GeV
 at $95\%$ confidence level, with down--type squarks further excluded up to
masses of $290$~GeV.

\end{abstract}

\vspace{1.5cm}

\begin{center}
Submitted to \EJC
\end{center}

\end{titlepage}

\begin{flushleft}

F.D.~Aaron$^{5,49}$,           
C.~Alexa$^{5}$,                
V.~Andreev$^{25}$,             
S.~Backovic$^{30}$,            
A.~Baghdasaryan$^{38}$,        
S.~Baghdasaryan$^{38}$,        
E.~Barrelet$^{29}$,            
W.~Bartel$^{11}$,              
K.~Begzsuren$^{35}$,           
A.~Belousov$^{25}$
% ...

\bigskip{\it
 $ ^{1}$ I. Physikalisches Institut der RWTH, Aachen, Germany \\
% ...
$ ^{5}$ National Institute for Physics and Nuclear Engineering (NIPNE) ,
          Bucharest, Romania$^{ m}$ \\
% ...
$ ^{11}$ DESY, Hamburg, Germany \\
% ...
$ ^{25}$ Lebedev Physical Institute, Moscow, Russia$^{ e}$ \\
% ...
$ ^{29}$ LPNHE, Universit\'e Pierre et Marie Curie Paris 6,
          Universit\'e Denis Diderot Paris 7, CNRS/IN2P3, Paris, France \\
 $ ^{30}$ Faculty of Science, University of Montenegro,
          Podgorica, Montenegro$^{ e}$ \\
% ...
$ ^{35}$ Institute of Physics and Technology of the Mongolian
          Academy of Sciences, Ulaanbaatar, Mongolia \\
% ...
$ ^{38}$ Yerevan Physics Institute, Yerevan, Armenia \\
% ...
$ ^{49}$ Also at Faculty of Physics, University of Bucharest,
          Bucharest, Romania \\
% ...
$ ^e$ Supported by the Deutsche Forschungsgemeinschaft \\
% ...
$ ^m$ Supported by the Romanian National Authority for Scientific Research
      under the contract PN 09370101 \\
}

\end{flushleft}

\newpage

\section{Introduction}

The $ep$ collider HERA is ideally suited to search for new particles coupling to 
electron\footnote{In the following the generic term {\it electron} refers to 
both electron and positron unless explicitly stated otherwise.
}--quark pairs.
In supersymmetric (SUSY) models with $R$--parity violation (\Rp ), squarks 
can couple to electrons and quarks via Yukawa couplings $\lambda'$.
At HERA, squarks 
with masses up to the electron--proton centre--of--mass energy, \mbox{$\sqrt{s}=319$~GeV}, 
could be produced resonantly via the fusion of the incoming 
$27.6$~GeV electron and a quark from the incoming $920$~GeV proton.
Squark decays typically result in a number of
high energetic particles in the final state, thus 
several complementary \mbox{multi--lepton} and \mbox{multi--jet} topologies 
are investigated.
The data used in this analysis
correspond  
to an integrated luminosity of 
\mbox{$255\,\mathrm{pb^{-1}}$} for $e^+p$ collisions and
\mbox{$183\,\mathrm{pb^{-1}}$} for $e^-p$ collisions
which represents the full data sample collected at \mbox{$\sqrt{s}=319$~GeV}. 
For the $e^-p$ sample, this represents an increase of a factor of thirteen compared to the
previous H1 analysis~\cite{Aktas:2004ij}, while for the $e^+p$ sample this corresponds to
a factor of four. 
The search presented here supersedes the results previously obtained by 
H1~\cite{Aktas:2004ij,PEREZ}.
Complementary direct searches for \Rp\  SUSY have been carried out at the LEP $e^+e^-$ 
collider~\cite{LEP,L3RPV}
and at the Tevatron $p\bar{p}$ collider~\cite{TEVATRON,D0RES}.
Indirect constraints from low energy precision observables are also available~\cite{BETA0NU,APV,CCU,barbier}.

\section{Phenomenology and Monte Carlo Simulation}
\subsection{Production of squarks in \boldmath{$R$}--parity violating supersymmetry}

\begin{figure}[t] 

  \begin{center}
   \epsfig{file=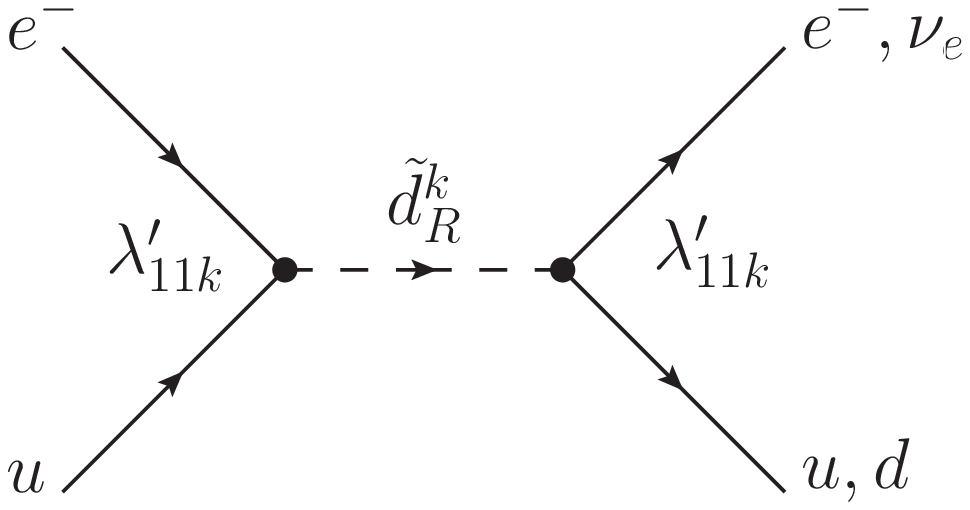,width=0.4\textwidth}
   \hspace{2cm}
   \epsfig{file=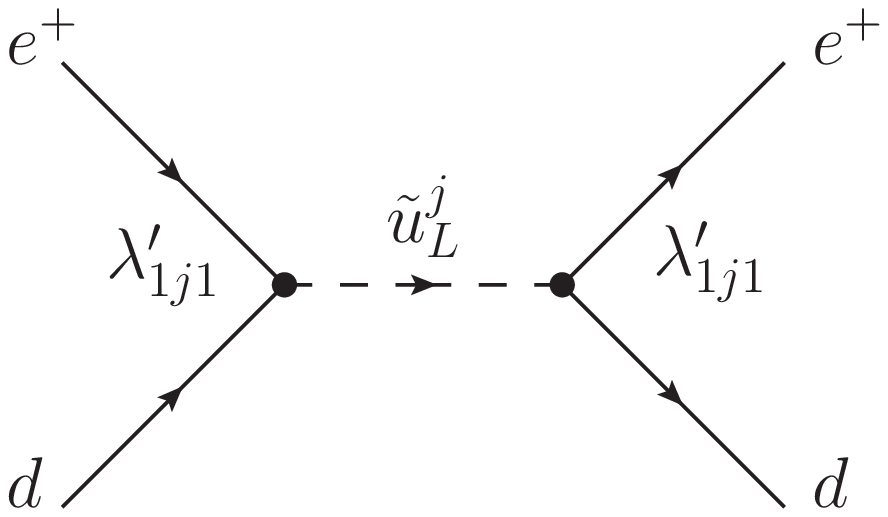,width=0.4\textwidth}
  \end{center}
  \begin{picture}(1,2)(1,2)
\put(20,5){(a)}
\put(107,5){(b)}
\end{picture}

  \caption{
  	Feynman diagrams for the single resonant s-channel production of right--handed
    down--type squarks in $e^-p$ collisions (a)  and left--handed up--type squarks in $e^+p$ collisions  (b)
    with subsequent decays into SM particles via Yukawa couplings $\lambda'_{11k}$ or 
    $\lambda'_{1j1}$, respectively. The right--handed down--type squarks 
can decay either into $e^- + {u}$ or $\nu_e + {d}$, while the left-handed up--type 
squarks decay into $e^+ + d$ only.
  	}
  \label{fig:directdecays}

\end{figure}

Supersymmetric extensions of the Standard Model (SM) introduce new elementary particles
which are the superpartners (sparticles) of SM particles but differ in spin by half a unit. 
A new quantum number $R_p=(-1)^{3B+L+2S}$ is defined, denoted $R$--parity, 
where $B$ is the baryon number, $L$ the lepton number 
and $S$ the spin of a particle. 
For particles $R_p=1$ and for their supersymmetric partners $R_p=-1$.
Most of the collider searches focus on SUSY models that conserve $R$--parity, allowing
only pair--production of sparticles. 
However, the most general supersymmetric theory that is renormalisable and gauge invariant 
with respect to the Standard Model 
gauge group does not impose $R$--parity 
conservation.
Couplings between two SM fermions and a squark ($\tilde{q}$) or a 
slepton ($\tilde{l}$) are then possible,
allowing the single production of sparticles.
The \Rp\ Yukawa couplings responsible for squark production at 
HERA originate from a lepton number violating term $\lambda'_{ijk} L_{i}Q_{j}\overline{D}_k$ 
in the superpotential, where $i,j$ and $k$ are family indices.
$L_i$, $Q_j$ and $\overline{D}_k$ are superfields, which contain the left--handed leptons, 
the left--handed up--type quarks and the right--handed down--type quarks, respectively, together 
with their SUSY partners.
Non--vanishing couplings $\lambda'_{1jk}$ allow the resonant production of 
squarks at HERA via $eq$ fusion~\cite{RPVIOLATION}.
Feynman diagrams of these processes are shown in figure~\ref{fig:directdecays}.
The values of the couplings are not fixed by the theory but are required to
be small 
to conform with present observations.
For simplicity, it is assumed here that one of the $\lambda'_{1jk}$ couplings
dominates over all the other
trilinear \Rp\ couplings.
At high Bjorken--$x$ the density of antiquarks in the 
proton is significantly smaller than that of the valence quarks. 
Hence $e^-p$ scattering gives sensitivity to the couplings $\lambda'_{11k}$ $(k=1,2,3)$
which dominate the production of $\tilde{d}_R$--type squarks (i.e. the superpartners $\tilde{d}_R$,
$\tilde{s}_R$ and $\tilde{b}_R$ of down--type quarks). The dominant contribution to the production
cross section is thus approximately proportional to $\lambda'^2_{11k} \cdot u(x)$ where $u(x)$ gives the probability to
find a $u$ quark in the proton carrying the momentum fraction $x=M^2_{\tilde{q}}/s$, where $M^{2}_{\tilde{q}}$ is
the squared mass of the produced squark.
By contrast, $e^+p$ scattering provides sensitivity to the couplings $\lambda'_{1j1}$ $(j=1,2,3)$
which dominate the production of $\tilde{u}_L$--type squarks (i.e. the superpartners $\tilde{u}_L$,
$\tilde{c}_L$ and $\tilde{t}_L$ of up--type quarks). Here the dominant contribution to the production
cross section is approximately proportional to $\lambda'^2_{1j1} \cdot d(x)$. Due to the
larger $u$ quark density in the proton at large $x$ with respect to the $d$ quark density,
larger production cross sections are expected in $e^-p$ interactions for identical couplings and squark masses.

Signal cross sections are obtained in the narrow width approximation 
by using the leading order amplitudes
given in~\cite{BUCHMULL}, 
corrected to account for next--to--leading order QCD effects using 
multiplicative correction factors~\cite{SPIRANLO}.
The parton densities are evaluated at the hard scale $M_{\tilde{q}}^2$.

\subsection{Final states from squark decays}

\begin{figure}[t] 
  \begin{center}
   \epsfig{file=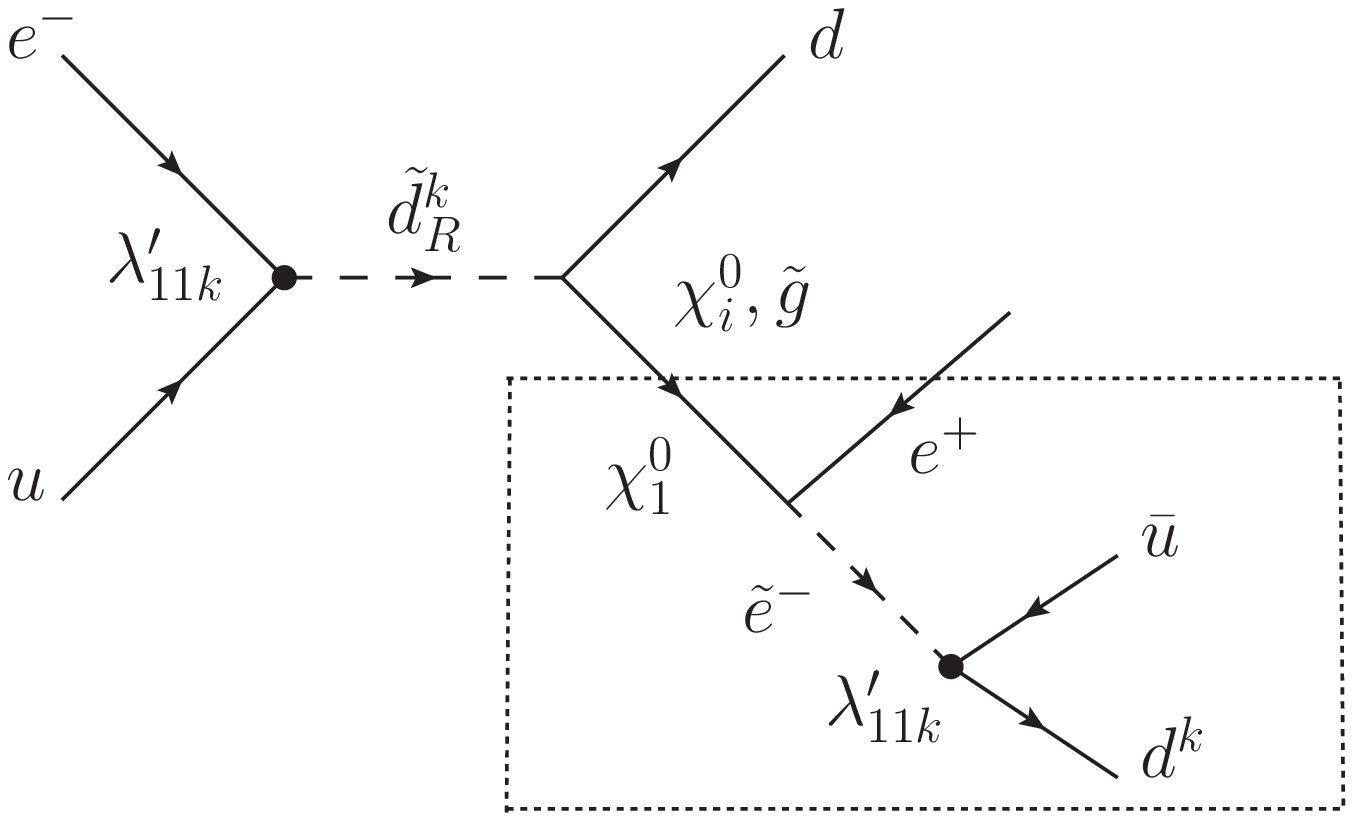,width=0.49\textwidth}
   \epsfig{file=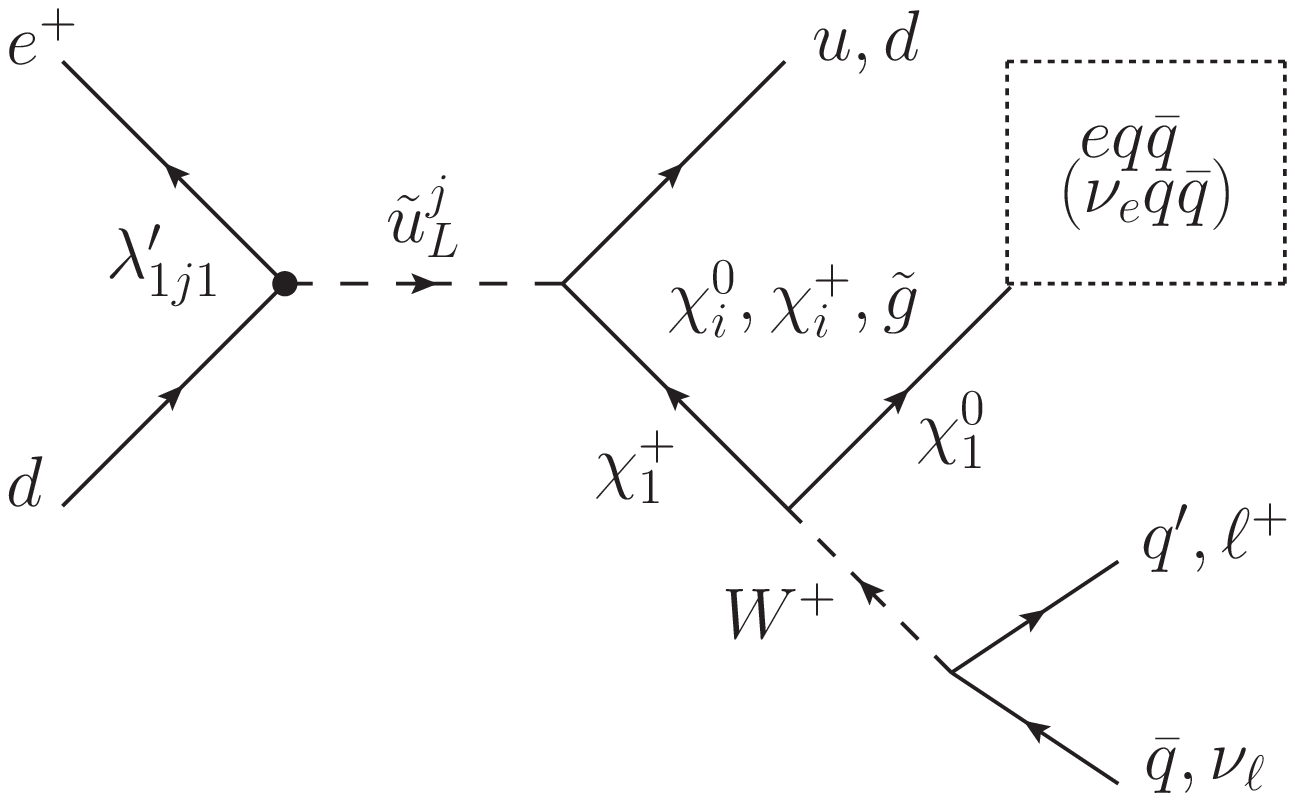,width=0.49\textwidth}
  \end{center}
    \begin{picture}(1,2)(1,2)
\put(20,5){(a)}
\put(107,5){(b)}
\end{picture}

  \caption{Feynman diagrams for squark decays proceeding via gauginos
  in the case of right--handed down--type  squarks (a)  and 
  left--handed up--type squarks (b) with subsequent 
    \Rp\ decay into SM fermions via Yukawa couplings $\lambda'_{11k}$ or 
    $\lambda'_{1j1}$, respectively. The resulting final states may 
    contain multi--leptons and multi--jets. The $\tilde{d}_R$--type squarks decay to 
    \mbox{$\chi^0_i$ ($i=1,2,3,4$)} or $\tilde{g}$, decays into charginos are suppressed while
    $\tilde{u}_L$--type squarks couple also to charginos $\chi^{+}_i$ ($i=1,2$).
}
  \label{fig:gaugedecays}
\end{figure}

In \Rp\ SUSY all sparticles are unstable.
Squarks can decay directly via the Yukawa coupling $\lambda'$ into SM fermions.
The $\tilde{d}^k_R$--type ($k=1,2,3$) squarks 
can decay via the coupling $\lambda'_{11k}$
either into $e^- + {u}$ or $\nu_e + {d}$, while the $\tilde{u}^j_L$--type ($j=1,2,3$) 
squarks decay via the coupling $\lambda'_{1j1}$
into $e^+ + d$ only,
as illustrated in figure~\ref{fig:directdecays}.
Squarks may also decay via $R_p$ conserving gauge couplings as illustrated in 
figure~\ref{fig:gaugedecays}.
The $\tilde{u}_L$--type squarks can undergo a gauge decay into states
involving a neutralino $\chi^0_i$ ($i=1,2,3,4$), a chargino $\chi^{+}_i$ ($i=1,2$) 
or a gluino $\tilde{g}$.
In contrast, $\tilde{d}_R$--type squarks mainly decay to $\chi^0_i$ or $\tilde{g}$ 
and decays into charginos are suppressed~\cite{RPVIOLATION}.
\par

 The squark decay chains analysed in this paper
 are classified by event topology\cite{PEREZ,Aktas:2004ij}.
 This classification relies on the number of isolated electrons, muons and 
 hadronic jets
 in the final state, and on the presence of missing energy (indicating
 undetected neutrinos).
 The channels labelled $eq$ and $\nu q$ are the squark decay modes 
 that proceed 
 directly via \Rp\ couplings 
 resulting in event topologies with an isolated electron or neutrino and a single jet. 
 The remaining channels result from the gauge 
 decays of the squark and are characterised by multijet ($M\!J$) final states with
 additional leptons.
 The channels 
 labelled $eM\!J$ and $\nu M\!J$ involve one or two gauginos 
 ($\chi$ or $\tilde{g}$) in the decay cascade.  In the $eM\!J$ channel $e^+$ and $e^-$ are possible in the final state,
 such that with respect to the incident lepton charge, a ``right'' (same sign) charge $eM\!J$(RC)
  and a ``wrong'' (opposite sign) charge $eM\!J$(WC) channel are distinguished.
 Channels with an electron and an additional charged lepton $\ell$ (where $\ell=e,\mu$) 
 denoted by $ee M\!J$, $e\mu M\!J$ or neutrinos and a 
 charged lepton $e \nu M\!J$, $\nu \mu M\!J$ (generically written as $e \ell M\!J$ and $\nu \ell M\!J$) 
 necessarily involve two gauginos.
 Decay patterns involving more than two gauginos are kinematically suppressed and are 
 therefore not explicitly studied here.
 Processes leading to final states with tau leptons are 
 not expected to increase the sensitivity of this analysis and are
 not explicitly investigated.
 A dedicated search for isolated tau leptons in H1 shows good agreement with the SM~\cite{Aaron:2009wp}.

\subsection{Event simulation}
\label{sec:massgeneration}\label{sec:MC}
 For each of the signal topologies described above a dedicated Monte Carlo (MC) simulation is
 done. For the direct lepton--quark decay channels $eq$ and $\nu q$, as shown in figure~\ref{fig:directdecays},
 events are generated using LEGO~\cite{LEGO}.
 For the gauge decays of squarks (figure~\ref{fig:gaugedecays})  events are generated using SUSYGEN3~\cite{SUSYGEN}.
 
\par
 To allow a model independent interpretation of the results, the squark decay processes 
 are simulated for a wide range of masses of 
the sparticles involved. 
The final states contain only SM fermions ($f$) considered as massless.
The squark mass is varied from $100$~GeV to $290$~GeV in steps of typically $25$~GeV. 
For gauge decays of squarks involving a gaugino, which decays directly via \Rp,
the process $\tilde{q} \rightarrow q \chi^0_1$ is generated for 
$\chi^0_1$ masses ranging between $30$~GeV  and $M_{\tilde{q}}$.
In order to study cascade gauge decays which involve two gauginos, the processes 
$\tilde{q} \rightarrow q \chi^+_1 \rightarrow q \chi^0_1 f \bar{f'}$ and
$\tilde{q} \rightarrow q \chi^0_2 \rightarrow q \chi^0_1 f \bar{f'}$ are generated for 
$\chi^+_1$  and $\chi^0_2$ masses ranging between $40$~GeV and $M_{\tilde{q}}$, and 
for $\chi^0_1$ masses between $30$~GeV and $M_{\chi^+_1}$ or $M_{\chi^0_2}$. 
The masses of the gauginos are varied in steps of approximately $10$~GeV.
The lower mass values for squarks and gauginos are motivated by the exclusion domains 
resulting from \Rp\ SUSY searches at LEP~\cite{LEP,L3RPV}.
The mass intervals are sufficiently 
small to allow linear interpolation
of signal detection efficiencies as a function 
of the masses of the sparticles involved.

 \par
 SM processes may mimic the characteristics of 
 the final states of squark decays.
 This SM background is dominated
 by neutral current (NC) and charged current (CC) deep inelastic scattering 
 (DIS), with additional small contributions from photoproduction, single $W$ boson production and lepton pair production.
 The RAPGAP~\cite{Jung:1993gf} event generator, which implements the Born level, QCD Compton and boson--gluon fusion
  matrix elements, is used to model inclusive NC DIS events. The QED radiative effects 
arising from real photon emission from both the incoming and outgoing electrons are simulated 
using the HERACLES~\cite{Kwiatkowski:1990es} program. Direct and resolved photoproduction of jets and prompt 
photon production are simulated using the PYTHIA~\cite{Sjostrand:2000wi} event generator. The simulation is based 
on Born level scattering matrix elements with radiative QED corrections. In RAPGAP and 
PYTHIA, jet production from higher order QCD radiation is simulated using leading logarithmic 
parton showers and hadronisation is modelled with 
Lund string fragmentation~\cite{lund}. 
Inclusive CC DIS events are simulated using the 
DJANGO~\cite{Schuler:yg} program, which includes first order leptonic QED radiative corrections based on 
HERACLES. The production of two or more jets in DJANGO is accounted for using the colour 
dipole model~\cite{Lonnblad:1992tz}. 
The leading order MC prediction of processes with 
two or more high transverse momentum jets in
NC DIS, CC DIS and photoproduction is scaled by a factor of $1.2$ to account for the incomplete description of 
higher orders in the MC generators~\cite{Adloff:2002au,Aktas:2004pz}. 
Contributions arising from the production of 
single W bosons and multi--lepton events are modelled using the EPVEC~\cite{Baur:1991pp} 
and GRAPE~\cite{Abe:2000cv} event generators, respectively. 
\par
Generated events are passed through a GEANT~\cite{Brun:1987ma} based simulation of the H1 apparatus, 
which takes into account the actual running conditions of the data taking. Simulated events are reconstructed 
and analysed using the same program chain as is used for the data.

\section{Experimental Method}
\subsection{H1 detector}
A detailed description of the H1 experiment can be found elsewhere~\cite{Abt:h1}.
Only the detector components relevant to this analysis are briefly described here.  
A right--handed cartesian  coordinate system is used with the origin at the nominal primary $ep$ interaction vertex. 
The proton beam direction defines the positive $z$ axis (forward direction). The polar angle $\theta$ and the transverse momenta $P_T$ of all particles 
are defined with respect to this axis.  The azimuthal angle $\phi$ defines the particle direction in the transverse plane. The pseudorapidity  is defined as $\eta=-\ln {\tan {\frac{\theta}{2}}}$.

The Liquid Argon (LAr) calorimeter~\cite{Andrieu:1993kh} covers the polar angle range
\mbox{$4^\circ < \theta < 154^\circ$} with full azimuthal acceptance.
The energies of electromagnetic showers are measured in the LAr with a precision of
\mbox{$\sigma (E)/E \simeq 11\%/ \sqrt{E/\mbox{GeV}} \oplus 1\%$} and hadronic energy depositions
with \mbox{$\sigma (E)/E \simeq 50\%/\sqrt{E/\mbox{GeV}} \oplus 2\%$}, as determined in test beam measurements~\cite{Andrieu:1993tz,Andrieu:1994yn}.
A lead--scintillating fibre calorimeter (SpaCal)~\cite{SpaCal} covering
  the backward region \mbox{$153^\circ < \theta < 178^\circ$} completes 
  the measurement of charged and neutral particles.
For electrons a relative energy resolution of 
\mbox{$\sigma (E)/E \simeq 7\%/\sqrt{E/\mbox{GeV}} \oplus 1\%$}
is reached, as determined in test beam measurements~\cite{SpaTestBeam}. 
The central \mbox{($20^\circ < \theta < 160^\circ$)}  and forward \mbox{($7^\circ < \theta < 25^\circ$)}  
inner tracking detectors are used to
measure charged particle trajectories and to reconstruct the interaction
vertex.
The LAr calorimeter and inner tracking detectors are enclosed in a super--conducting magnetic
coil with a field strength of $1.16$~T.
From the curvature of charged particle trajectories in the magnetic field, the central tracking system provides transverse momentum measurements with a resolution of 
\mbox{$\sigma_{P_T}/P_T = 0.005 P_T / \mbox{GeV} \oplus 0.015$}~\cite{Kleinwort:2006zz}.
The return yoke of the magnetic coil is the outermost part of the detector and is
equipped with streamer tubes forming the central muon detector
\mbox{($4^\circ < \theta < 171^\circ$)}. 
In the very forward region of the detector \mbox{($3^\circ < \theta < 17^\circ$)} 
a set of drift chambers detects muons and measures their momenta using an iron toroidal magnet.
The luminosity is determined from the rate of the Bethe--Heitler process $ep \rightarrow ep \gamma$,
measured using a photon detector located close to the beam pipe at $z=-103~{\rm m}$, in the backward direction.

\subsection{Particle identification and event reconstruction}

Electromagnetic particle (electron and photon) candidates are identified as compact and isolated clusters of energy
in the electromagnetic part of the LAr calorimeter. 
Electron candidates are identified as electromagnetic particle candidates with 
an associated track.
Identification of
muon candidates is based on a track in the inner tracking detectors, associated to a signal in the muon system.
Tracks and calorimeter deposits not identified as originating from isolated 
electromagnetic particles or muons are combined into
cluster--track objects
to reconstruct the hadronic final state~\cite{benji}.
Jets are reconstructed from these objects 
using an inclusive $k_T$ algorithm~\cite{kT1,kT2} with a minimum $P_T$ of $4$~GeV. The missing transverse momentum 
\mbox{${P}_T^{\rm{miss}}$}, which may indicate the presence of neutrinos in the final state, is
derived from all reconstructed particles in the event.
A neutrino four--vector $P_{\nu}$ is reconstructed by
exploiting momentum and energy conservation.
The transverse momentum of the neutrino is reconstructed by assuming one
neutrino with significant energy in the event $ \vec{P}_T^{\nu} \equiv \vec{P}_T^{\rm{miss}} $.
The energy of the neutrino is then reconstructed 
exploiting the energy and longitudinal momentum balance:
$   \sum_{i} \left( E^i-P^i_z \right) + \left( E^{\nu} - P^{\nu}_z \right) 
   = 2E^{0}_e = 55.2\,\mbox{GeV}, 
$ 
where the sum runs over all detected particles, $P_z$ is the momentum along the proton
beam axis and $E_e^0$ denotes the energy of the incident electron.

For further selection the following observables are used which in SM DIS events correspond to the Lorentz--invariant quantities:
inelasticity $y$, negative four--momentum transfer squared $Q^2$ and Bjorken's scaling variable 
$x$. They can be reconstructed as:
$$y_e = 1 - \frac{E_e (1 - \cos \theta_e) }{2E_e^0},
\;\;\;Q^2_e = \frac{P^2_{T,e}}{1-y_e},\;\;\;x_e = \frac{Q^2_e} {y_e s},$$
where the polar angle $\theta_e$, energy 
$E_e$ and transverse momentum $P_{T}^{e}$ of the electron with the highest 
$P_T$ found in the event are used.
If no electron is reconstructed in the event similar quantities can be calculated using the Jacquet--Blondel method~\cite{BLONDEL}
from the hadronic final state variables:
$$y_h=\frac{\sum \left(E-P_z\right)_h}{2E_e^0},
\;\;\;Q^2_h= \frac{P^2_{T,h}}{1-y_h},\;\;\;x_h = \frac{Q^2_h} {y_h s},$$
where $P_{T,h}$ is the transverse momentum of the hadronic final system 
calculated from the before mentioned cluster--track objects.
The sum ${\sum \left(E-P_z\right)_h}$ runs over energies and momenta of jets only.

\subsection{Trigger and data quality}
All data events used for this search 
are
 triggered by the LAr calorimeter~\cite{Adloff:2003uh}.  
Events with an electromagnetic deposit in the LAr with an energy greater than $10$~GeV are detected 
with an efficiency close to $100\%$~\cite{nikiforov}.
Events are also triggered by hadronic jets, with a trigger efficiency above $95\%$ 
for a jet transverse momentum $P_T^{\rm{jet}} > 20$~GeV 
and almost $100\%$ for $P_T^{\rm{jet}} > 25$~GeV~\cite{matti}.
For events with missing transverse energy of $20$~GeV, the trigger efficiency 
is about $90\%$ and increases to above $95\%$ for missing transverse 
energy above $30$~GeV~\cite{trinh}.
\par
In order to remove background events induced by cosmic showers and other non--$ep$ sources, 
the event vertex is required to be within $35$~cm in $z$ of the 
mean position for $ep$ collisions.
In addition, topological filters and timing vetoes are applied~\cite{negri} in order to reject beam related and cosmic background.

\section{Data Analysis}
The event selection is carried out in several exclusive analysis channels.
The resulting
event rates for each channel and the range of 
efficiencies for the selection of signal events 
are given in table~\ref{tab:totnum}.
\subsection{Electron--jet final state \boldmath{$eq$}}
The final state of a squark decaying into 
an electron and a high $P_T$ jet is identical to the NC DIS signature at high $x$ and $Q^2$.
For the signal the reconstructed invariant mass distribution \mbox{$M_e= \sqrt{x_e s}$} shows a resonance peak at the nominal squark mass with a resolution
\mbox{$\delta M_e = 4-10$~GeV} depending on the mass of the squark.
Differences in the $M_e$ and $y_e$ distributions of the two processes 
allow to discriminate them statistically.
Squarks produced in the $s$--channel decay isotropically leading to a flat $d\sigma/dy$ distribution,
whereas for NC DIS events a distribution proportional to $1/y^{2}$ is expected.
\par
Events are selected by requiring \mbox{${P}_T^{\rm{miss}}<15$~GeV} and
\mbox{$40$~GeV $< \sum (E-P_z) < 70$~GeV} where the sum runs over all particles in the final
state.  
An isolated electron with \mbox{$P_{T}^{e}>16$~GeV} 
in the region \mbox{$5^{\circ}<\theta_e<145^{\circ}$}, \mbox{$Q^2_e>2500$~GeV$^2$} and \mbox{$y_e<0.9$} is required. 
The high $y$ region is excluded to reduce background contributions arising from photoproduction events. 
An $M_e$ dependent cut on $y_e$ is determined by minimising the expected limit
from signal and SM MC. 
The cut ranges from \mbox{$y_e>0.5$} for
masses around $100$~GeV to \mbox{$y_e>0.2$} for masses around $290$~GeV.
In order to remain exclusive with respect to other channels, events with 
muon candidates with \mbox{$P_T^{\mu}>5$~GeV} or
two jets with \mbox{$P_{T}^{\rm{jet}}>15$~GeV} are rejected.
\par
\begin{figure}[t] 
  \begin{center}
    \epsfig{file=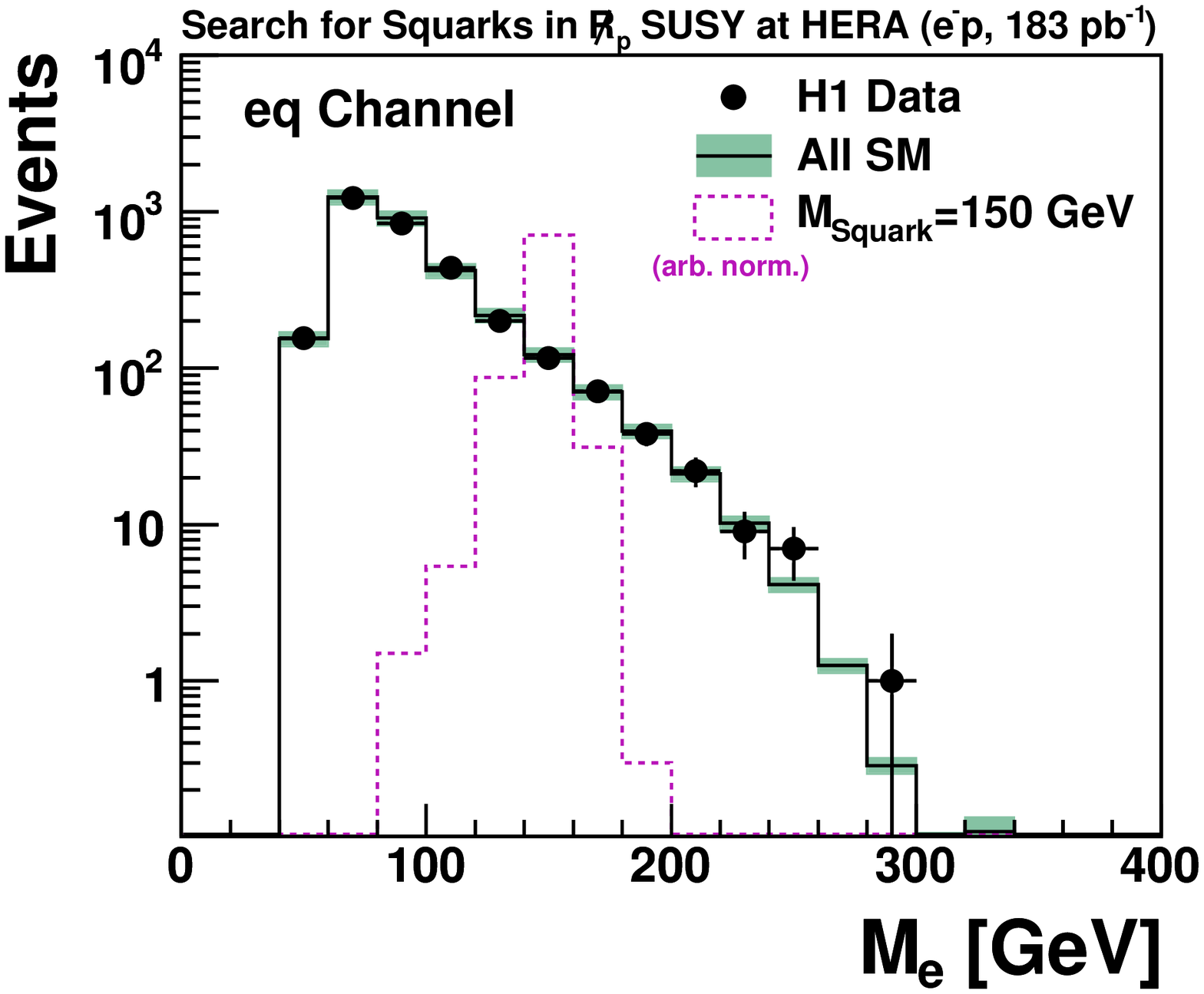,width=0.45\textwidth}
    \epsfig{file=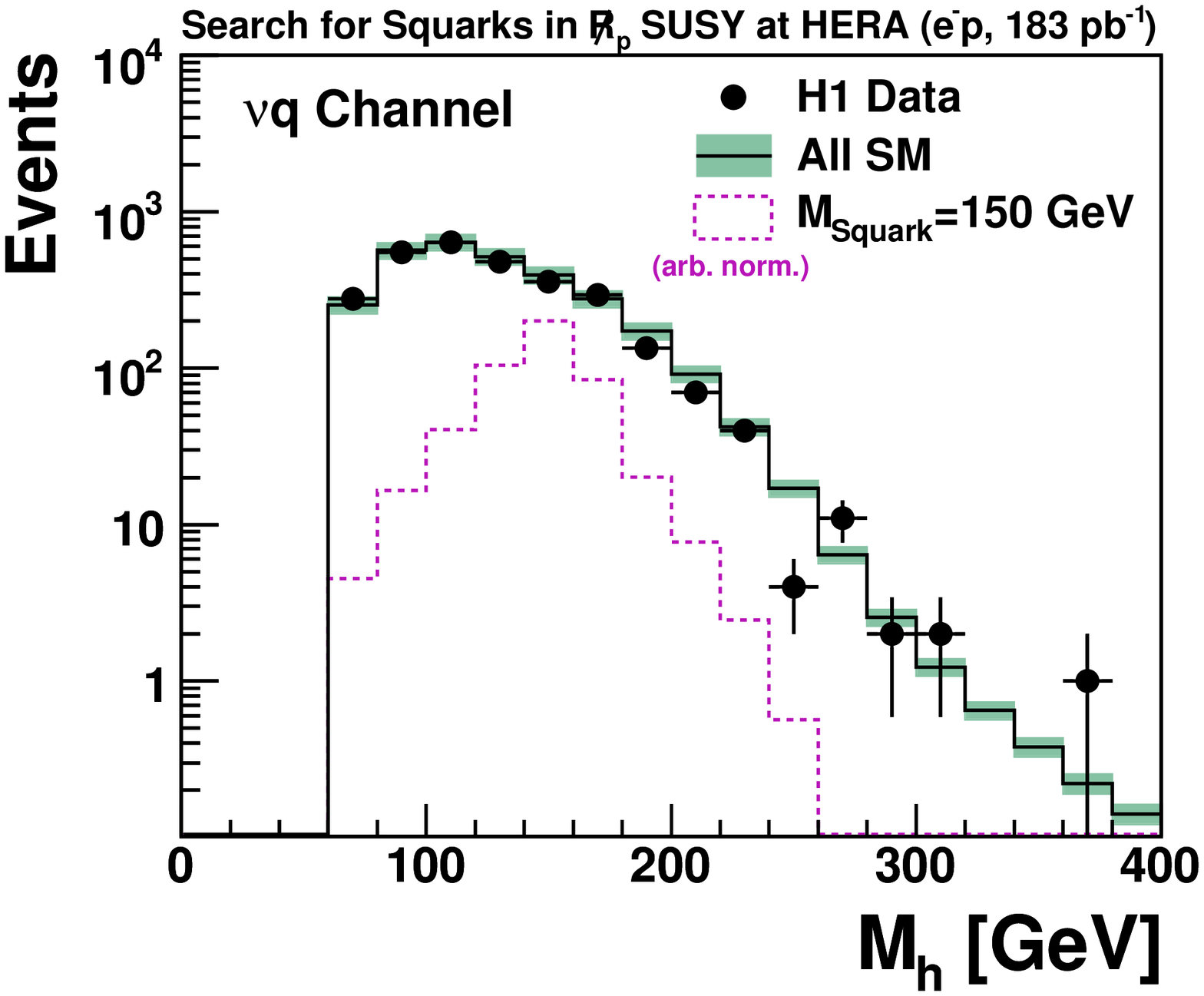,width=0.45\textwidth}\\
    \epsfig{file=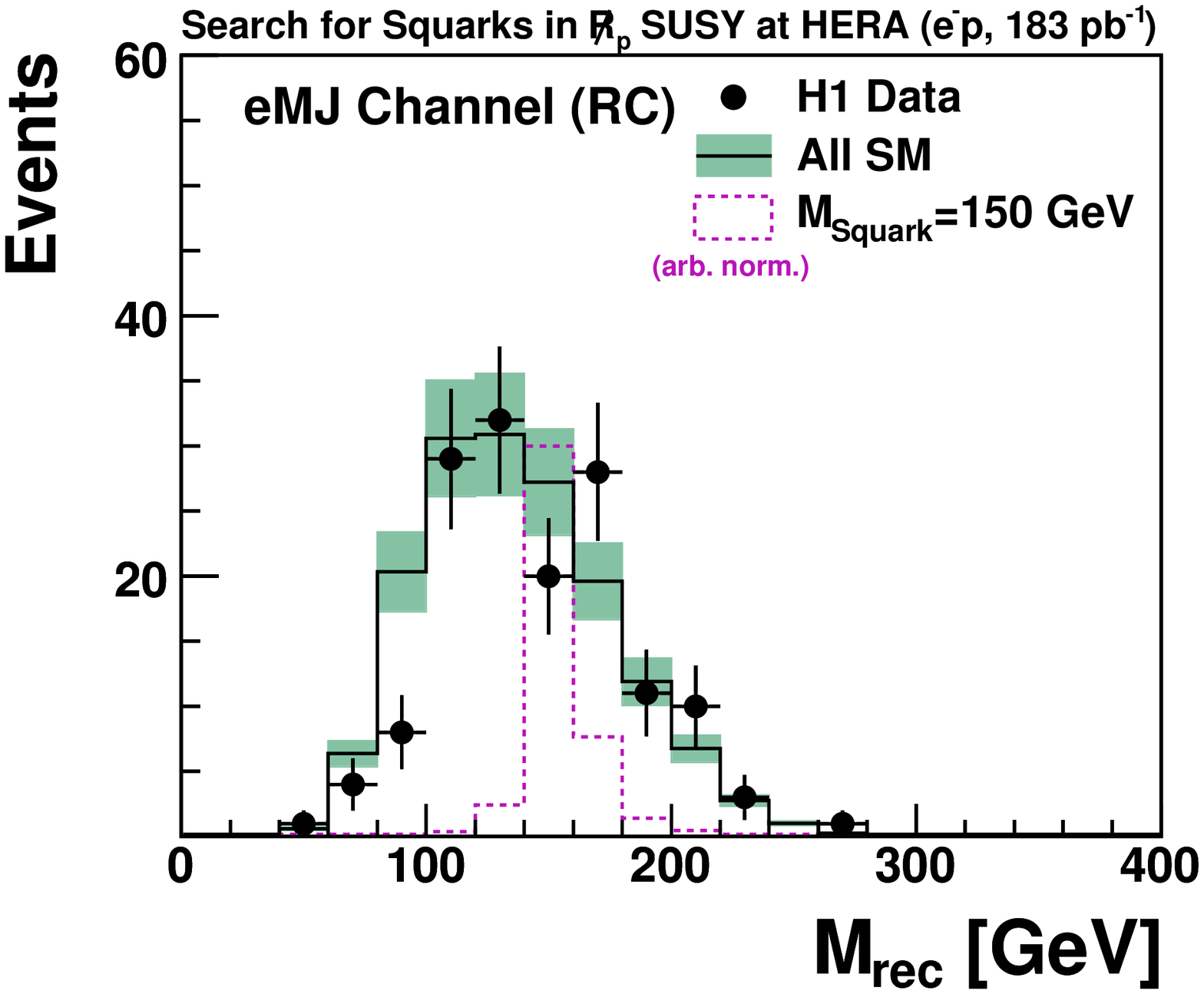,width=0.45\textwidth}
   \epsfig{file=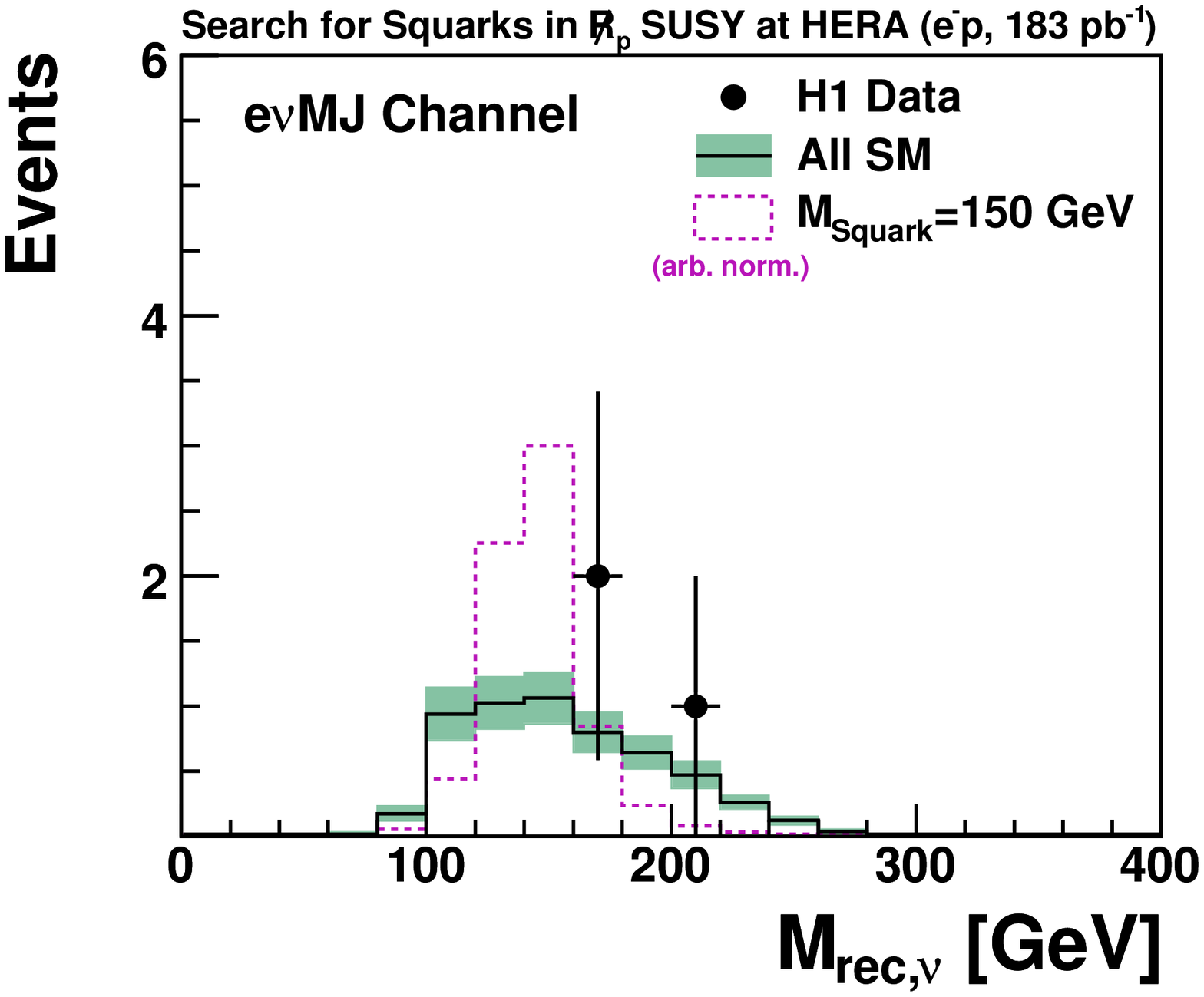,width=0.45\textwidth}
   \epsfig{file=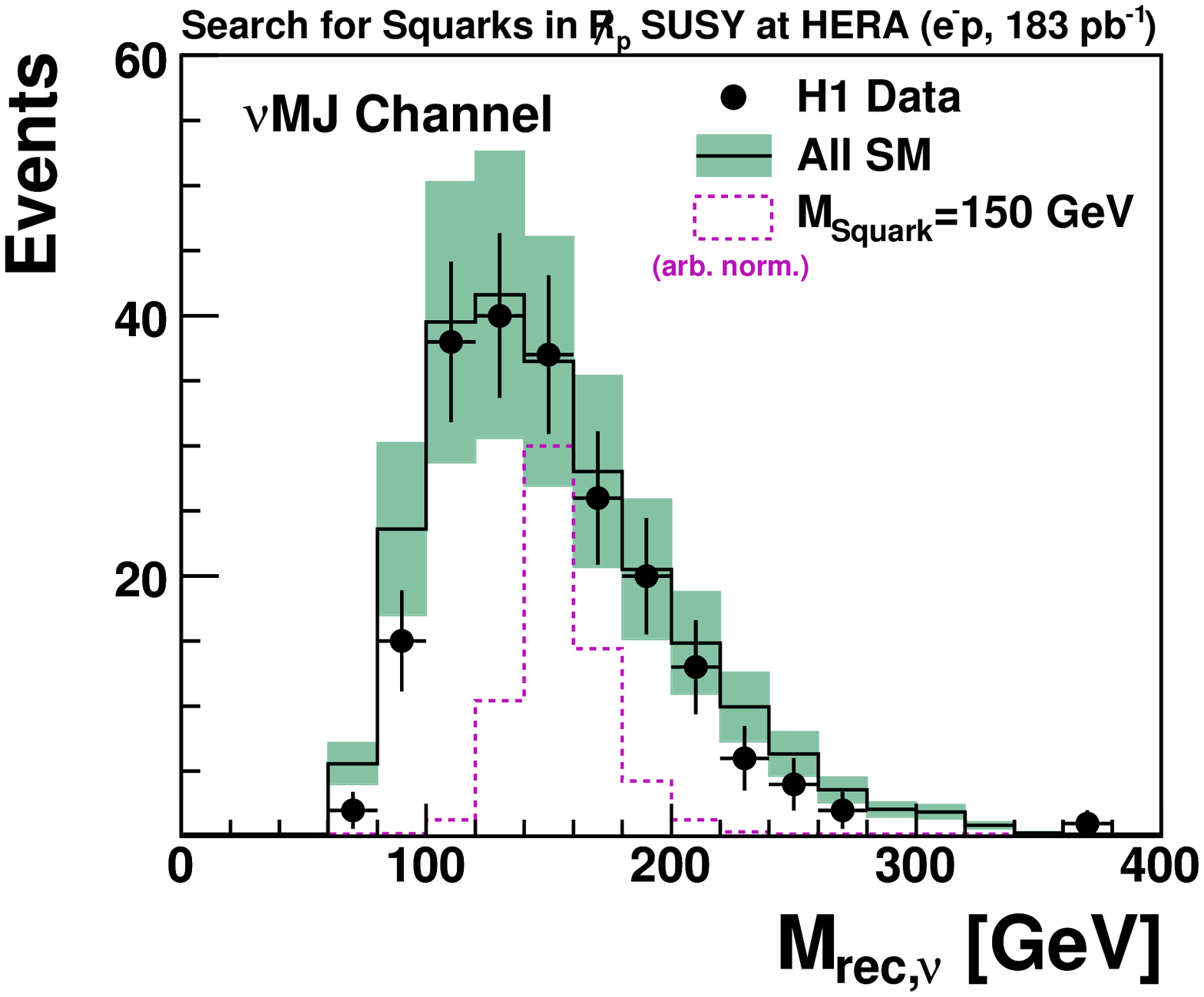,width=0.45\textwidth}
  \end{center}
  \begin{picture}(1,2)(1,2)
\put(67,170){(a)}
\put(140,170){(b)}
\put(67,110){(c)}
\put(140,110){(d)}
\put(103,47){(e)}
\end{picture}
  \caption{Reconstructed invariant mass distributions in all selection channels with 
  data (points) events from $183$~pb$^{-1}$ of $e^{-}p$ 
  collisions compared to SM MC predictions. 
    The method used for the reconstruction ($M_e,M_h,M_{\rm{rec}},M_{\rm{rec},\nu}$)
  depends on the analysis channel.
  The error band gives all model and experimental systematic
uncertainties on the SM prediction (solid histogram) added in quadrature.
Error bars of data points show statistical uncertainties.
  The dashed histogram indicates the signal from a squark with 
  $M_{\tilde q} = 150\,\text{GeV}$ with arbitrary normalisation.}
  \label{fig:masselec}

\end{figure}

The $M_e$ spectra after this selection for data and SM background are shown in figures~\ref{fig:masselec}a 
and~\ref{fig:massposi}a for $e^-p$ data and $e^{+}p$ data, respectively.
For both data samples no significant deviation from the SM expectation is observed. 
In the $e^{-}p$ sample, 
$3121$
events are observed while the SM expectation yields 
\mbox{$3215\pm 336$}.
In the $e^+p$ data sample 
$2946$
candidate events are found compared to
\mbox{$2899\pm 302$}
expected from SM processes. 

\subsection{Neutrino--jet final state \boldmath{$\nu q$}}
Squarks decaying into a neutrino and a high $P_T$ jet lead to the same signature as CC DIS events with
high missing transverse momentum. Similarly to the $eq$ channel, the resonant 
$s$--channel production and isotropic decay allows a statistical separation of 
signal and background. The resolution $\delta M_h$ of the reconstructed 
invariant mass $M_h= \sqrt{x_h s}$ varies between \mbox{$12$ and $22$~GeV} depending on the mass of the squark considered.
\par
The presence of a neutrino in the event is required by imposing
\mbox{${P}_T^{\rm{miss}}>30$~GeV} and\linebreak
\mbox{$\sum (E-P_z) < 50$~GeV}.
The phase space is restricted to $Q^2_h>2500$~GeV$^2$ and $y_h<0.9$.
Similar to the $eq$ channel, a cut on $y_h$  dependent on the reconstructed 
mass $M_h$ is applied. The cut ranges from \mbox{$y_h>0.3$} for
masses around $100$~GeV to \mbox{$y_h>0.1$} for masses around $290$~GeV.   
In order to remain exclusive with respect to other channels, events with any electron or 
muon candidate with \mbox{$P_T^{e, \mu}>5$~GeV} or events 
containing two jets with \mbox{$P_{T}^{\rm{jet}}>15$~GeV} are rejected.
\par
Only $\tilde{d}_R$--type squarks, which are produced dominantly in $e^{-}p$ collisions, can undergo 
direct decay leading to a $\nu q$ final state. 
For selected events the $M_h$ spectrum of this data set and of the 
simulation of SM background events is shown in figure~\ref{fig:masselec}b. No
significant deviation from the SM expectation is found. 
In data $2858$
events are observed while 
\mbox{$2983\pm 358$} 
are expected from SM processes. 

\subsection{Electron--multijet and  electron--lepton--multijet final states}

\subsubsection*{Common preselection for \boldmath{$eM\!J(RC),eM\!J(WC),eeM\!J,e\mu M\!J$ and $e\nu M\!J$}}
Squarks decaying via neutralinos or charginos are expected to have a higher multiplicity of jets and leptons in the
final state.
 The signatures correspond
to final states detectable in higher order NC DIS processes.
However, as heavy particles are boosted forward,
  in events with squarks 
the decay products are mainly
emitted into the forward part of the detector. 
This feature is used to distinguish between the signal and SM background.
\par
A common preselection is applied for the $eM\!J(RC),eM\!J(WC),eeM\!J,e\mu M\!J$ and $e \nu  M\!J$ channels:
At least one isolated electron with \mbox{$P_{T}^{e}>6$~GeV} and \mbox{$E_e>11$~GeV} in the region \mbox{$5^{\circ}<\theta_e<110^{\circ}$} is required. 
In addition, the condition $y_e>0.3$ is used to reduce the background from NC
DIS.
Central electrons with \mbox{$30^\circ<\theta_e<110^\circ$} are required to have a well measured track associated to the cluster and
the distance of closest approach between the track impact point and the centre--of--gravity of the cluster should not exceed $12$~cm.
Furthermore
the energy of the track is required to match the energy of the associated cluster according to \mbox{$E_{{\rm{cluster}}}/E_{{\rm{track}}}>0.5$}.
At least two jets with \mbox{$P_{T}^{\rm{jet}}>15$~GeV} in the jet polar angle range 
\mbox{$7^{\circ}<\theta_{\rm{jet}}<145^{\circ}$} are also required.
\par
By requiring \mbox{$Q^2_e> 1000$~GeV$^2$} the steep decrease of the NC DIS cross section with increasing $Q^2$ is exploited.
This corresponds to an implicit upper cut on the electron polar angle as $Q^2$ is strongly
correlated with the polar angle of the scattered electron.
For signal events at least one high $P_T$ particle is expected to be emitted in the forward direction,
therefore the highest $P_T$ electron or one of the two highest $P_T$ jets 
has to fulfill \mbox{$\theta_{e,{\rm{jet}}}<40^{\circ}$}.
Moreover, of the two highest $P_T$ jets, the one with the largest polar angle, 
$\theta_{\rm{backw}}$, must satisfy the condition 
\mbox{$\theta_{{\rm{backw}}} <  (y_e - 0.3) \cdot 180^{\circ} $}, separating efficiently 
signal events from NC DIS background~\cite{JOHTHESIS}.

For selected events an invariant mass $M_{\rm{rec}}$ is calculated as \mbox{$M_{\rm{rec}}= \sqrt{4E^0_e (\sum{E_i}-E^0_e)}$}, where the energies $E_i$ of electrons, 
muons and jets found in the event
with \mbox{$P_T^{\rm{jet}}>5$~GeV} are included in the sum. The resolution 
$\delta M_{\rm{rec}}$ of this method ranges between \mbox{$6$ and $10$~GeV} depending on the mass of the squark considered.

\subsubsection*{Electron--multijet final state \boldmath{$eM\!J(RC),eM\!J(WC)$}}
Decays of squarks via gauginos are likely to produce a single isolated electron and multiple jets
in the final state.
Events produced in $e^{\pm}p$ collisions with exactly one electron
 and multiple jets in the final state are also
expected in the SM, where in general the measured charge of the electron corresponds to
that of the incident electron. A selection channel labelled ``right'' (same sign) charge \mbox{$eM\!J $ (RC)}
is used for events fulfilling this criterion. 
A selection channel where the
electron charge is identified as opposite to the incident electron, denoted
``wrong'' (opposite sign) charge \mbox{$eM\!J $ (WC)}, represents therefore a powerful test of the SM and is expected to be essentially
background free.
The distinction between RC and WC $eM\!J $ events is based on the 
curvature of the electron track measured in the central tracking system.
Events are allocated to the WC channel if the electron is found in the central region \mbox{$30^{\circ} <\theta_e<110^{\circ}$}
and its charge is measured to be opposite to that of the incident electron, with a charge significance greater than two standard deviations~\cite{JOHTHESIS}.  Otherwise the event is assigned to the RC channel.

Events are selected from the common preselection described above
by requiring in addition \mbox{$P_{T}^{\rm{miss}} < 15$~GeV} and
\mbox{$40$~GeV $ < \sum (E - P_{z}) < 70$~GeV} since no neutrinos are expected in these
channels. To ensure that the selection channels are exclusive, no additional
electron or muon candidate with \mbox{$P_{T}^{e, \mu} > 5$~GeV} may be present in the
event.

\par
\begin{figure}[t] 
  \begin{center}
    \epsfig{file=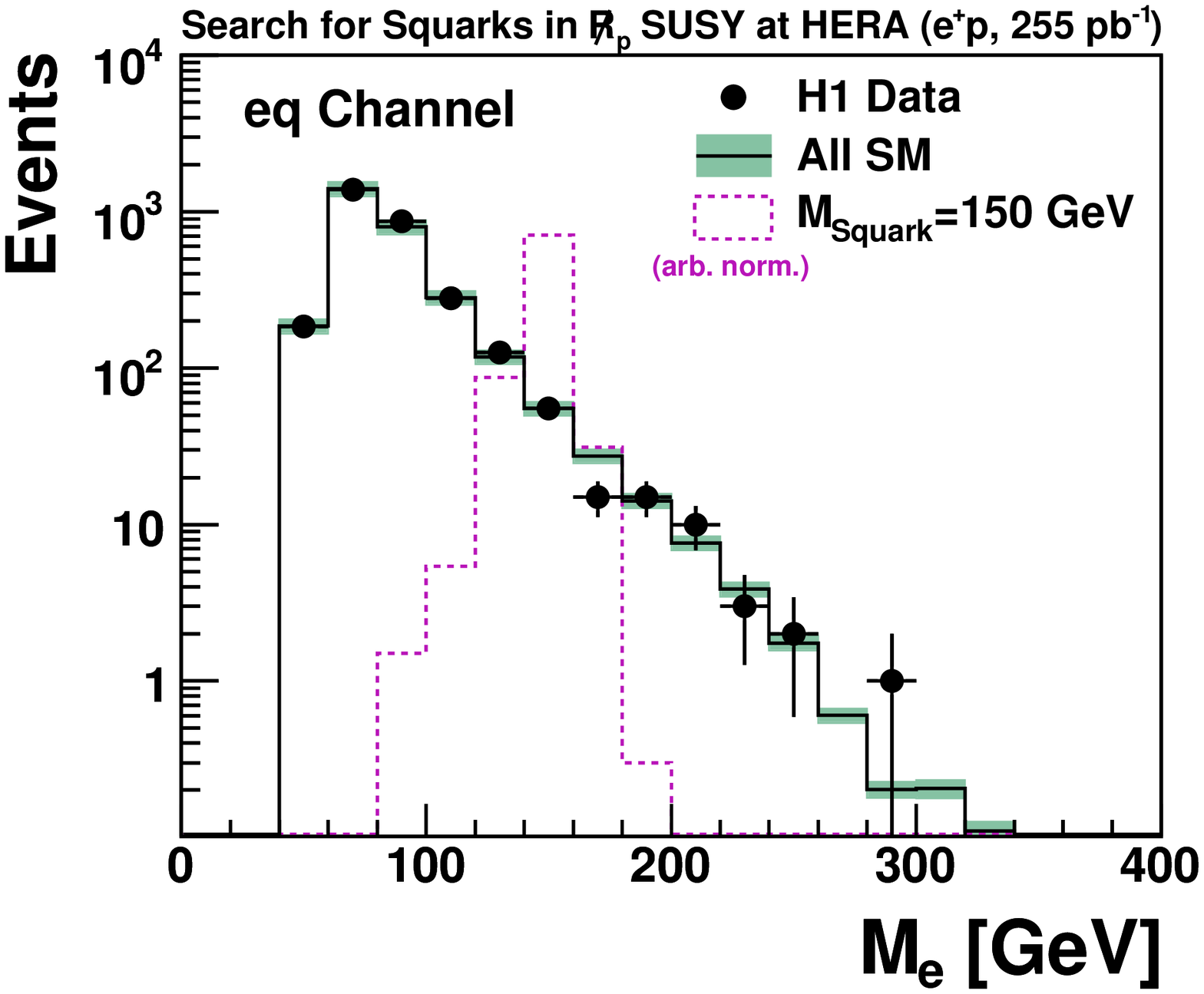,width=0.45\textwidth} 
    \epsfig{file=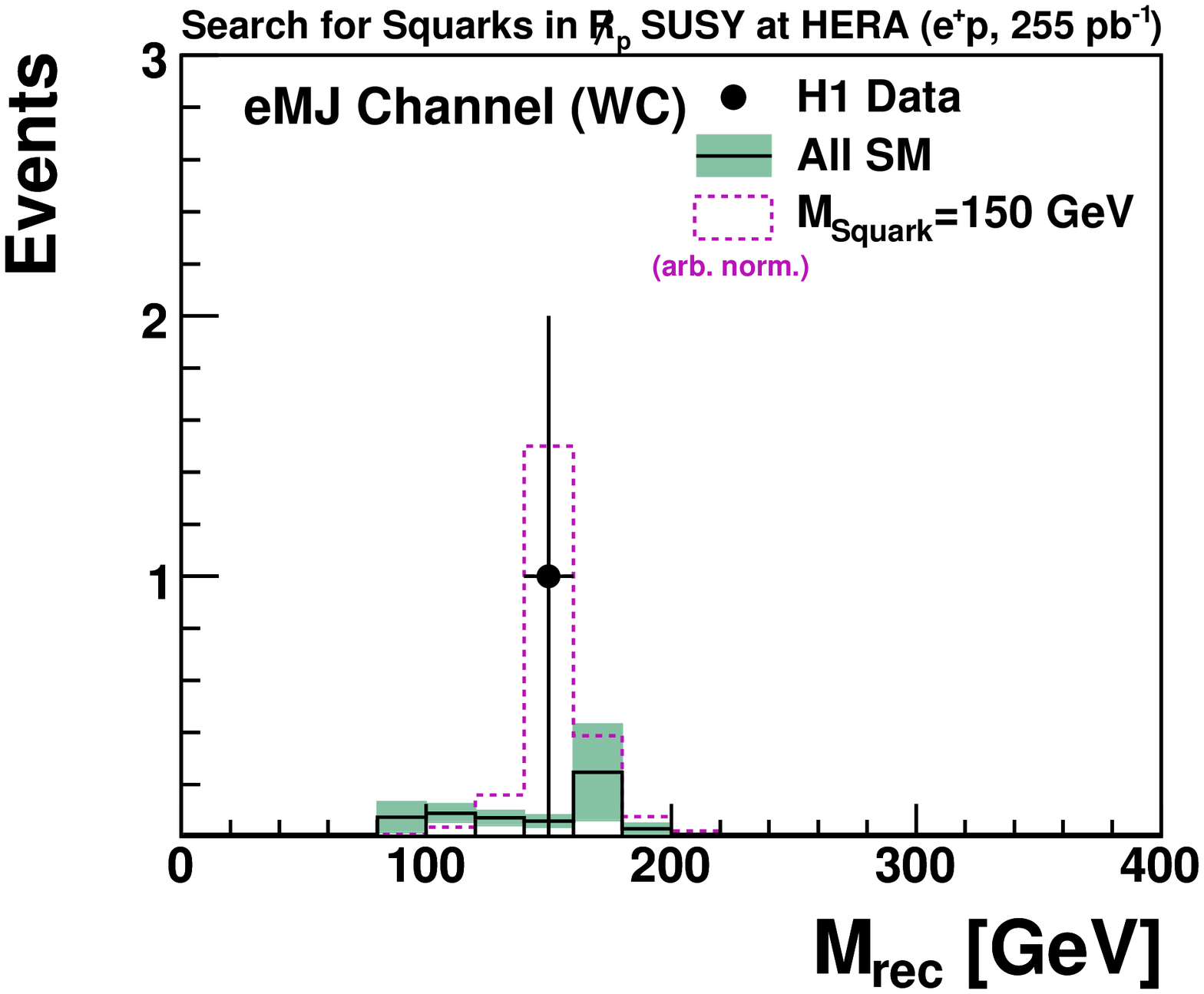,width=0.45\textwidth}\\
    \epsfig{file=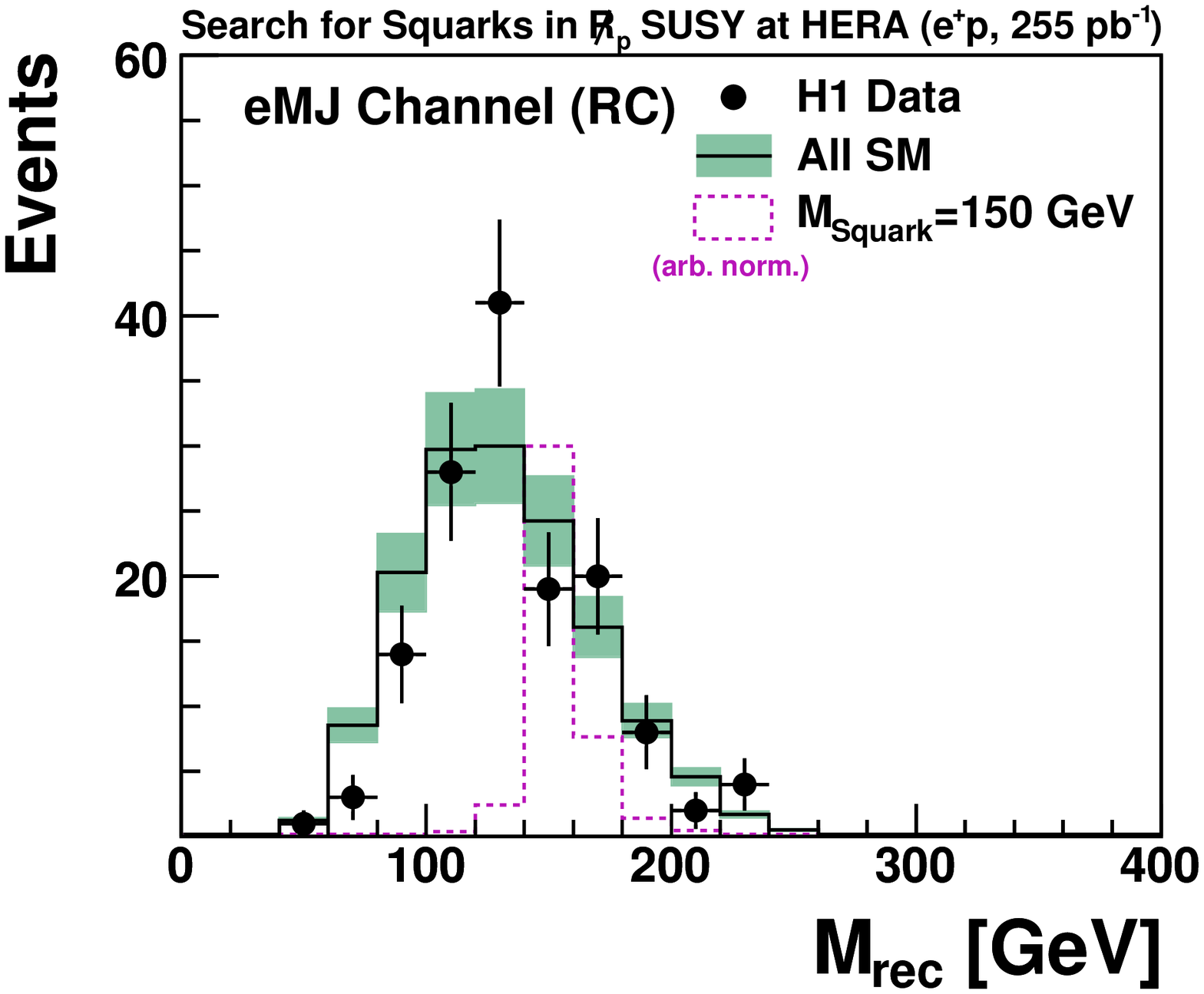,width=0.45\textwidth}
   \epsfig{file=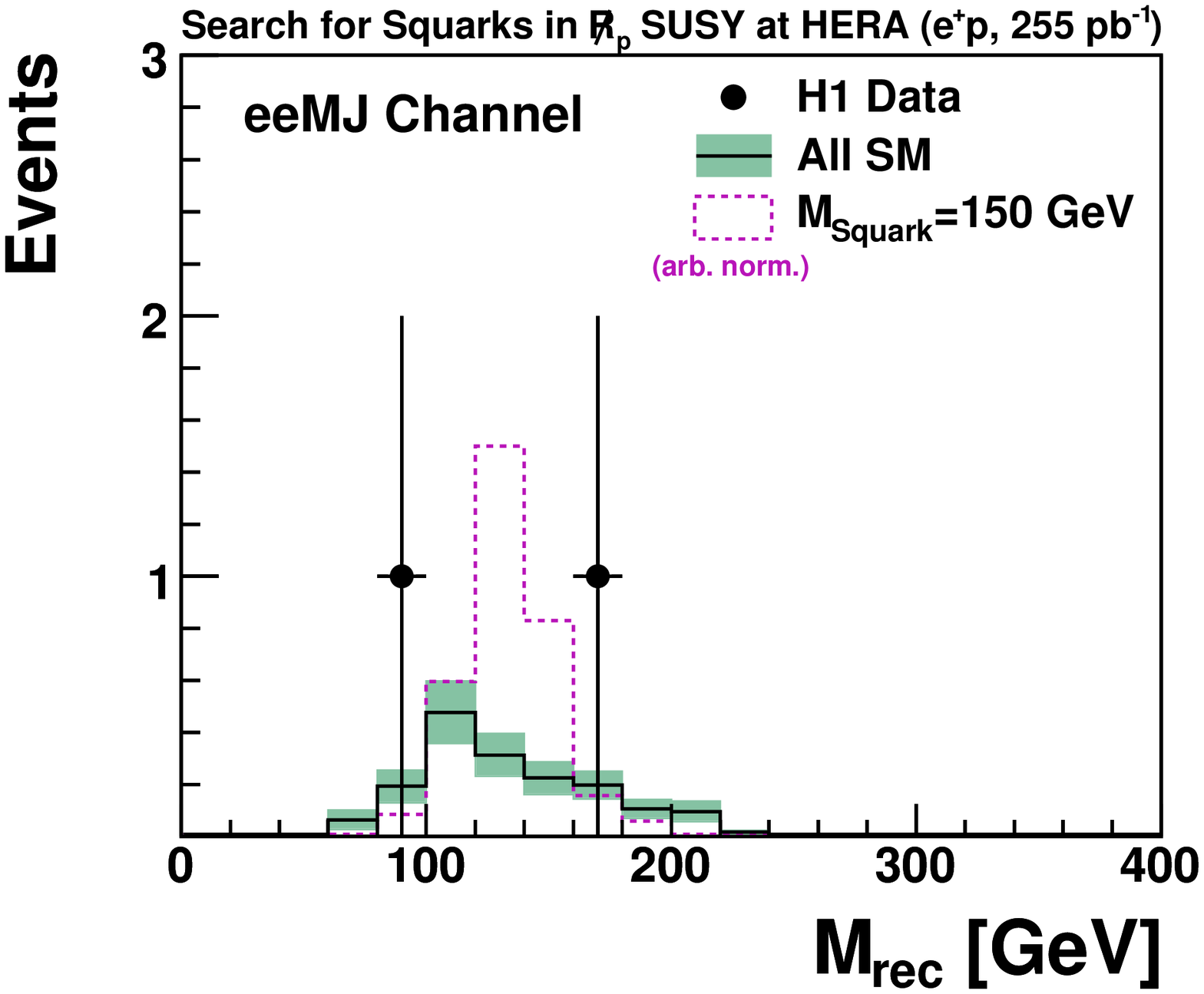,width=0.45\textwidth}
   \epsfig{file=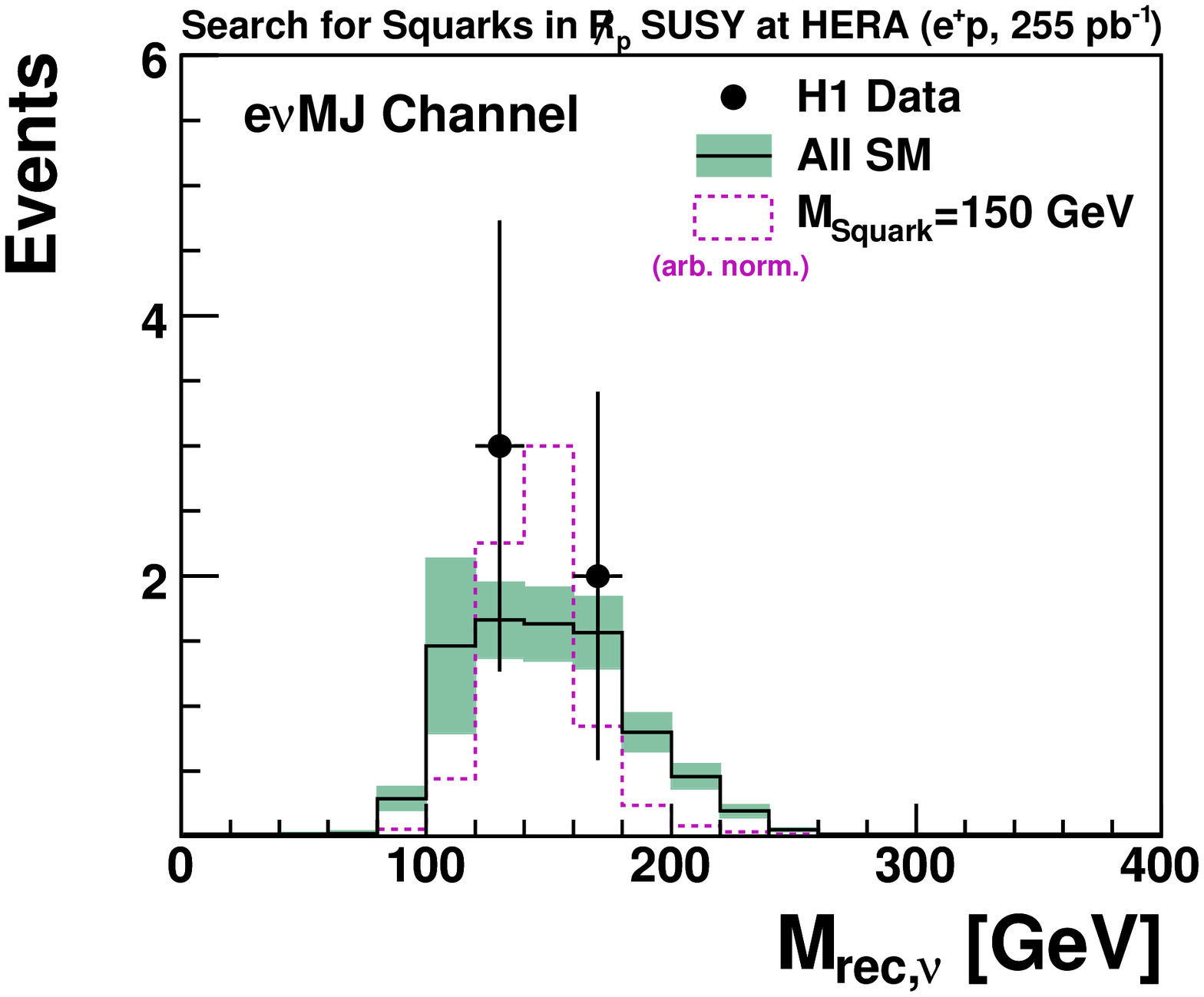,width=0.45\textwidth}
      \epsfig{file=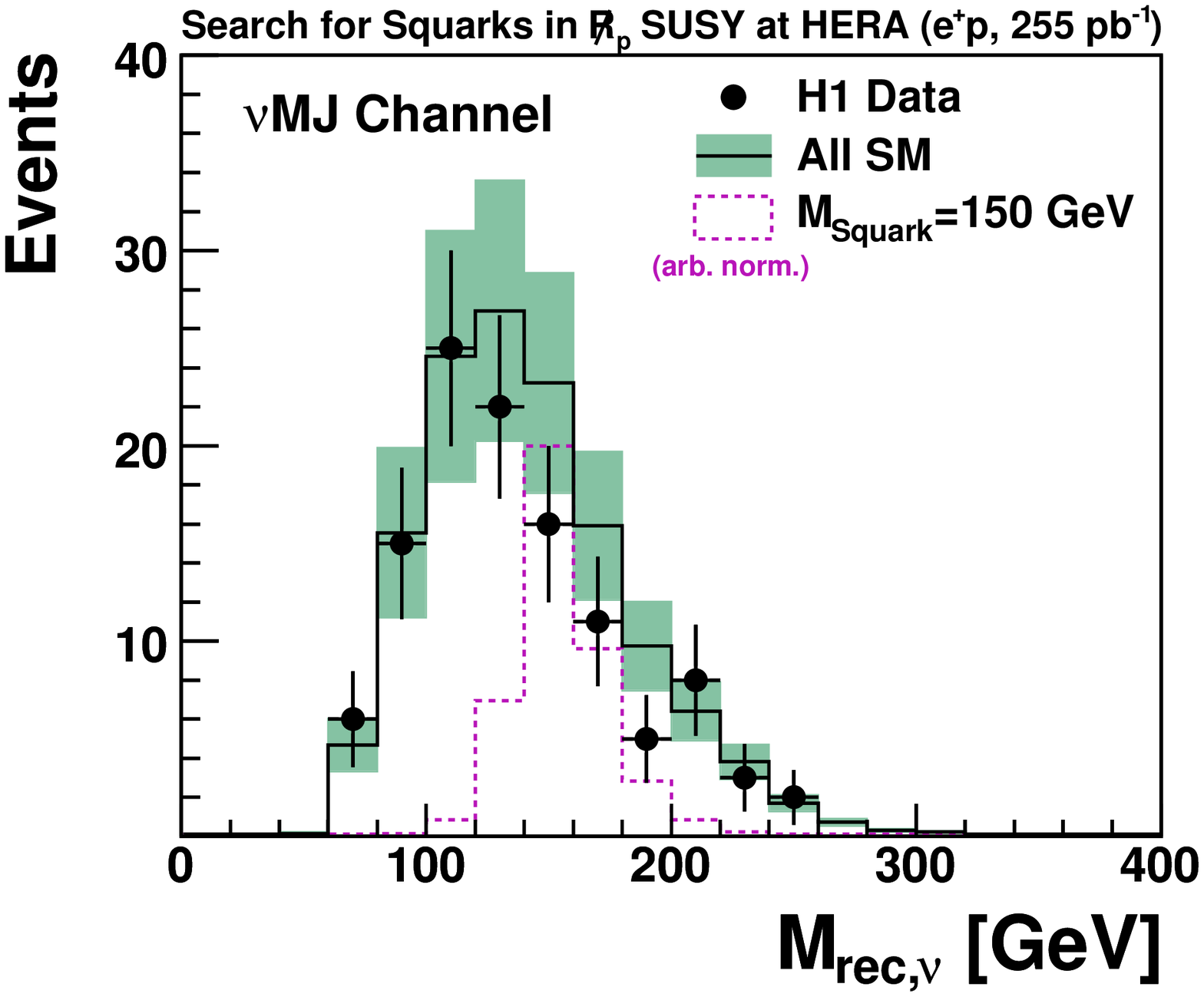,width=0.45\textwidth}

  \end{center}
\begin{picture}(1,2)(1,2)
\put(67,170){(a)}
\put(140,170){(b)}
\put(67,110){(c)}
\put(140,110){(d)}
\put(67,47){(e)}
\put(140,47){(f)}
\end{picture}
  \caption{Reconstructed invariant mass distributions in all selection channels with 
  data (points) events from $255$~pb$^{-1}$ of $e^{+}p$ 
  collisions compared to SM MC predictions. 
  The method used for the reconstruction ($M_e,M_{\rm{rec}},M_{\rm{rec},\nu}$)
  depends on the analysis channel.
  The error band gives all model and experimental systematic
uncertainties on the SM prediction (solid histogram) added in quadrature.
Error bars of data points show statistical uncertainties.
  The dashed histogram indicates the signal from a squark with 
  $M_{\tilde q} = 150\,\text{GeV}$ with arbitrary normalisation.}
  \label{fig:massposi}

\end{figure}

In the \mbox{$eM\!J $ (WC)} channel no event is observed in the $e^-p$ data while 
\mbox{$1.3\pm 0.3$}
 are expected
from SM processes and one candidate event is observed in the $e^+p$ data while 
\mbox{$0.6\pm 0.4$}
 are expected. For $e^+p$ collisions the mass spectrum is shown in figure~\ref{fig:massposi}b.
 \par

For events in the \mbox{$eM\!J $ (RC)} channel
an $M_{\rm{rec}}$ dependent cut on $y_e$ is applied to increase the sensitivity for signal events. 
The cut ranges from \mbox{$y_e>0.7$} for
masses around $100$~GeV to \mbox{$y_e>0.5$} for masses around $290$~GeV.   
The $M_{\rm{rec}}$ distributions for data and simulation are shown in figures~\ref{fig:masselec}c 
and~\ref{fig:massposi}c for $e^-p$ and $e^{+}p$ collisions, respectively. 
No significant deviation from the SM expectation is observed. 
In total 
$147$ events are observed in the $e^-p$ data while the SM simulation yields 
\mbox{$158.3\pm 23.9$}
and in the $e^+p$ data 
$140$ 
events are observed for an expectation of 
\mbox{$146.0\pm 21.4$}
 from SM processes. 
 
\subsubsection*{Electron--lepton--multijet final states, \boldmath{$eeM\!J,e\mu M\!J,e \nu M\!J$}}
Final states from squark decays may contain more than one isolated lepton if
the decays proceed via cascades of gauginos.
In addition to the common preselection
the $eeM\!J$ and $e\mu M\!J$ channels require  either an 
additional electron with the same criteria as described in the common preselection
or an isolated muon with \mbox{$P_{T}^{\mu}>5$~GeV} in the polar angle range \mbox{$10^{\circ}<\theta_{\mu}<110^{\circ}$}.  
After applying this selection
the SM background expectation is 
 very low in these channels.
For the \mbox{$e\mu M\!J$} channel there are no candidate events observed in the
$e^-p$ and $e^+p$ collision data for a SM expectation of \mbox{$0.03\pm 0.02$}
and \mbox{$0.03\pm 0.03$}, respectively. 
In the $e^-p$ collision data
 no candidate event for a SM expectation of 
\mbox{$1.5\pm 0.5$} is observed in the $eeM\!J$ channel
and two events in the $e^+p$ data are observed compared to 
 a SM expectation of \mbox{$1.7\pm 0.5$}.
The mass spectrum for the $eeM\!J$ channel in the $e^+p$ data
 is shown in figure~\ref{fig:massposi}d.
\par
In the $ e\nu M\!J$ channel a neutrino is expected in the final state, 
therefore in addition to the common preselection, large missing transverse 
momentum \mbox{${P}_T^{\rm{miss}}>15$~GeV} is required.
Due to the presence of the neutrino \mbox{$\sum (E-P_z)$} is significantly
reduced causing $y_h$ to be substantially smaller than $y_e$, while in 
NC DIS events  $y_h\approx y_e$ is expected.
Thus a cut \mbox{$y_e(y_e-y_h)>0.04$} is used to discriminate the SUSY signal 
from background events~\cite{JOHTHESIS}.  
Exclusivity with respect to the $eeM\!J$ and $e\mu M\!J$ 
channels is achieved
by rejecting events containing an additional electron or muon with \mbox{$P_T^{e, \mu}>5$~GeV}. 
The method used to reconstruct a mass $M_{\rm{rec}, \nu}$ for selected events
taking the energy of the neutrino into account is
explained in section~\ref{sec:vmj}. 

In the $e^-p$ collision data 
three
events are observed in the $ e\nu M\!J$ channel while 
\mbox{$5.6\pm 1.2$}
are expected and in $e^+p$ collision data 
five
events are observed while 
\mbox{$8.2\pm 2.0$}
are expected from SM processes. 
The mass spectra are shown in figures~\ref{fig:masselec}d 
and~\ref{fig:massposi}e for the $e^-p$ data and $e^{+}p$ data, respectively. 

\par
\subsection{Neutrino--multijet and neutrino--muon--multijet final states}
\label{sec:vmj}
\subsubsection*{Common preselection for \boldmath{$\nu M\!J$} and \boldmath{$\nu \mu M\!J$}}
Squark decays with single or multiple neutrinos produced via neutralino or chargino decays
can result in final states similar to that of higher order CC DIS processes.  

A substantial missing transverse momentum \mbox{${P}_T^{\rm{miss}}>26$~GeV} is required and
at least two jets must be found with \mbox{$P_{T}^{\rm{jet}}>15$~GeV} in the range \mbox{$7^{\circ}<\theta_{\rm{jet}}<145^{\circ}$}.
No electron candidate with \mbox{$P_T^{e}>5$~GeV} is allowed.
 A cut \mbox{$\sum (E - P_{z})<50$~GeV} is used to ensure the neutrino energy is positive.

For each selected event a squark mass $M_{\rm{rec},\nu}$ is calculated 
as \mbox{$M_{\rm{rec}, \nu}= \sqrt{4E^0_e (\sum{E_i}-E^0_e)}$} 
where the sum includes the energies of the
reconstructed neutrino, electrons, muons and jets with \mbox{$P_T^{\rm{jet}}>5$~GeV} in the event.
This method assumes that all missing energy is carried by a single neutrino
and yields a resolution $\delta M_{\rm{rec}, \nu}$
of about \mbox{$15$ to $20$~GeV} depending on the squark mass.

\subsubsection*{Neutrino--multijet final state \boldmath{$\nu M\!J$}}
Squark decays via gauginos are likely to produce final states with multiple jets and
a single neutrino.
Events are selected in the $\nu M\!J$ channel if no muon candidate is found. 
\par
A cut on $y_h$ dependent on the reconstructed mass $M_{\rm{rec}, \nu}$
is applied to enhance the signal.
The cut ranges from \mbox{$y_h>0.5$} for
masses around $100$~GeV to \mbox{$y_h>0.4$} for masses around $290$~GeV.   
The $M_{\rm{rec},\nu}$ spectra are shown in figures~\ref{fig:masselec}e 
and~\ref{fig:massposi}f for $e^-p$ and $e^{+}p$ collision data and SM background 
simulation. 
In the $e^-p$ data 
$204$
candidate events are selected while 
\mbox{$235.5\pm 63.3$}
are expected and in the $e^+p$ data 
$113$
candidate events are selected while 
\mbox{$134.0 \pm 33.8$}
are expected from SM processes.

\subsubsection*{Neutrino--muon--multijet final state \boldmath{$\nu \mu M\!J$}}
If  an isolated muon with \mbox{$P_{T}^{\mu}>5$~GeV} in the polar angle range \mbox{$10^{\circ}<\theta_{\mu}<110^{\circ}$} is found
in an event in the common $\nu M\!J$ selection, the event is
classified as $\nu \mu M\!J$ candidate. 
No candidate events are found,  in the $e^-p$ or in the $e^+p$ collision data, in agreement with the SM expectations of
\mbox{$0.04\pm 0.02$}
and \mbox{$0.06\pm 0.03$}, respectively. 

\subsection{Systematic uncertainties}
The following experimental systematic uncertainties are considered:
\begin{itemize}
\item The uncertainty on the electromagnetic  energy scale varies depending on 
the polar angle from $0.7\%$ in the central region to $2\%$ in the forward region~\cite{trinh}. The polar angle measurement  uncertainty of electromagnetic clusters is $3$~mrad. 
\item The jet  energy scale is known within $2\%$~\cite{trinh}. The uncertainty on the jet polar angle determination is $10$ mrad.
\item The luminosity measurement has an uncertainty of $3\%$.
\end{itemize}
The effects of the experimental systematic uncertainties on the SM expectation and signal efficiencies are determined by
varying the corresponding experimental quantities 
within one standard deviation
in the MC samples
and propagating the variations 
to the final distributions.
The resulting experimental uncertainties are determined for each analysis channel individually 
and added in quadrature.
In the $eq$ channel the uncertainty 
on the overall SM event yield
was found to be $3\%$, while in the ${\nu}q$ channel an uncertainty of
$7\%$ is determined. In the ${e}M\!J$ (RC and WC) channels the  
resulting systematic uncertainty amounts to $4\%$, in the ${e}{\mu}M\!J$ and
${e\nu}M\!J$ channels to $7\%$, in the ${ee}M\!J$ channel 
to $10\%$ and in the ${\nu}M\!J$ and ${\nu}{\mu}M\!J$ channels to $20\%$.

Additional model uncertainties are attributed to the SM MC event generators described in
section~$\ref{sec:MC}$. A conservative error of $10\%$ is attributed to NC (RAPGAP) and CC (DJANGO) DIS processes with
only one high $P_T$ jet. 
To account for the uncertainty on higher order QCD corrections, an
uncertainty of $15\%$ is attributed to NC DIS and photoproduction processes (PYTHIA) with at least two high $P_T$ jets.
The normalisation uncertainty of CC DIS processes with at least two high $P_T$ jets is estimated to be $20\%$~\cite{trinh}. 
A $5\%$ uncertainty is attributed to the contribution from multi--lepton events (GRAPE) and a $15\%$ uncertainty on the production of single W bosons (EPVEC).
These uncertainties include contributions from the proton parton distribution functions and from missing
higher order QCD corrections. 
The total error on the SM prediction is determined by adding the effects of all model and experimental systematic
uncertainties in quadrature.

For the signal cross section further uncertainties arise from the determination of signal efficiencies
($10\%$ due to available MC statistics), 
the theoretical uncertainty on the squark production cross section ($7\%$ for low squark masses,
up to $50\%$ for the highest masses from the PDF uncertainty) and an uncertainty due to the scale at
which the PDFs are evaluated ($7\%$)~\cite{JOHTHESIS}.

\section{Exclusion Limits}

No significant deviation from the SM expectation is observed in any channel.
Consequently the observations in all analysis channels are 
combined to set constraints on various supersymmetric models.
Exclusion limits are obtained on
the production of squarks parameterised by the strength of the \Rp\ couplings $\lambda'_{1j1}$ and $\lambda'_{11k}$ and
dependent on the mass of the squark.

\subsection{Procedure}

For the interpretation of the results a version of the Minimal Supersymmetric 
Standard Model (MSSM) is considered
where the masses of the neutralinos, charginos and gluinos
are determined via the usual parameters:
the ``Higgs--mass'' term $\mu$, which mixes the Higgs superfields; 
the SUSY soft--breaking mass parameter $M_2$;
and the ratio of the vacuum expectation values of the two neutral scalar Higgs fields $\tan \beta$~\cite{MSSM}.
The parameters are defined at the electroweak scale.

A set of parameters $(\tan \beta,\mu,M_2)$ together with the sfermion masses 
and a coupling $\lambda'_{1jk}$ define a supersymmetric 
scenario where the masses of the gauginos and the branching
ratios for squark decays into the different final state topologies are 
fixed and can be obtained using the SUSYGEN3~\cite{SUSYGEN} package. 
The branching ratios for the specific parameters of the model are taken into 
account in the combination~\cite{Junk:1999kv}. A sliding mass window technique is used in channels with high 
contributions of irreducible SM background 
($eq$, $\nu q$, $eM\!J$ and $\nu M\!J$)
to improve the signal to background ratio for the squark mass examined. 
The width of the mass window is determined by minimising the expected limit,
and increases towards high squark masses, reflecting 
the corresponding mass reconstruction resolution. 
The small efficiency losses due to the finite mass window width are taken
into account.
A $95\%$ confidence level (CL) upper limit $\sigma_{\rm lim}$ on 
the squark production cross section compatible with the simultaneous observation in all channels is derived using a
modified frequentist approach based on Likelihood ratios\cite{Junk:1999kv}. Sets of model parameters
leading to signal cross sections above $\sigma_{\rm lim}$ are excluded.

If the squark width is non--negligible, in particular for squark masses 
approaching the kinematic limit,
the production cross section decreases at the resonance peak and contributions 
from the lower tail of the squark mass distribution become important, enhanced
by the rapid increase of proton parton distributions at low Bjorken--$x$~\cite{JOHTHESIS}.
This is taken into account by generating events for negligible squark widths 
to determine signal efficiencies at all masses. The selection
efficiencies are then corrected for the actual squark width by reducing 
the efficiency for the signal selection accordingly~\cite{JOHTHESIS}.

In the special case of stop and sbottom squark production, namely via \Rp\ couplings $\lambda'_{131}$ and $\lambda'_{113}$,
the mixing of the weak eigenstates \mbox{$\tilde{t}_L,\tilde{t}_R$} (and \mbox{$\tilde{b}_L,\tilde{b}_R$}) to the mass eigenstates 
 \mbox{$\tilde{t}_1,\tilde{t}_2$ ($\tilde{b}_1,\tilde{b}_2$)} via an angle $\theta_{\tilde{t}}$ ($\theta_{\tilde{b}}$)
becomes important for the calculation of branching ratios and production cross sections. 
The gauge decay via a top quark would lead to
decay products different from the first two generation squarks, for which the 
efficiencies are determined. 
Final states with top quarks are not considered explicitly.
Since top signals 
would in any case be present in one of the selection
topologies, this approach is conservative.

\subsection{Constraints on a phenomenological MSSM}
Constraints are set in a scenario of a phenomenological MSSM~\cite{MSSM}
where the lightest supersymmetric particle (LSP) is the neutralino $\chi^0_1$.
Slepton masses $M_{\tilde{l}}$ are fixed at $90$~GeV, close to the lowest 
mass bound from \Rp\ sfermion searches at LEP~\cite{LEP} and  
squark masses are treated as free parameters.
For higher slepton masses only very small degradations in the derived
constraints are expected~\cite{Aktas:2004ij}. 

\begin{figure}[t] 
  \begin{center}
    \epsfig{file=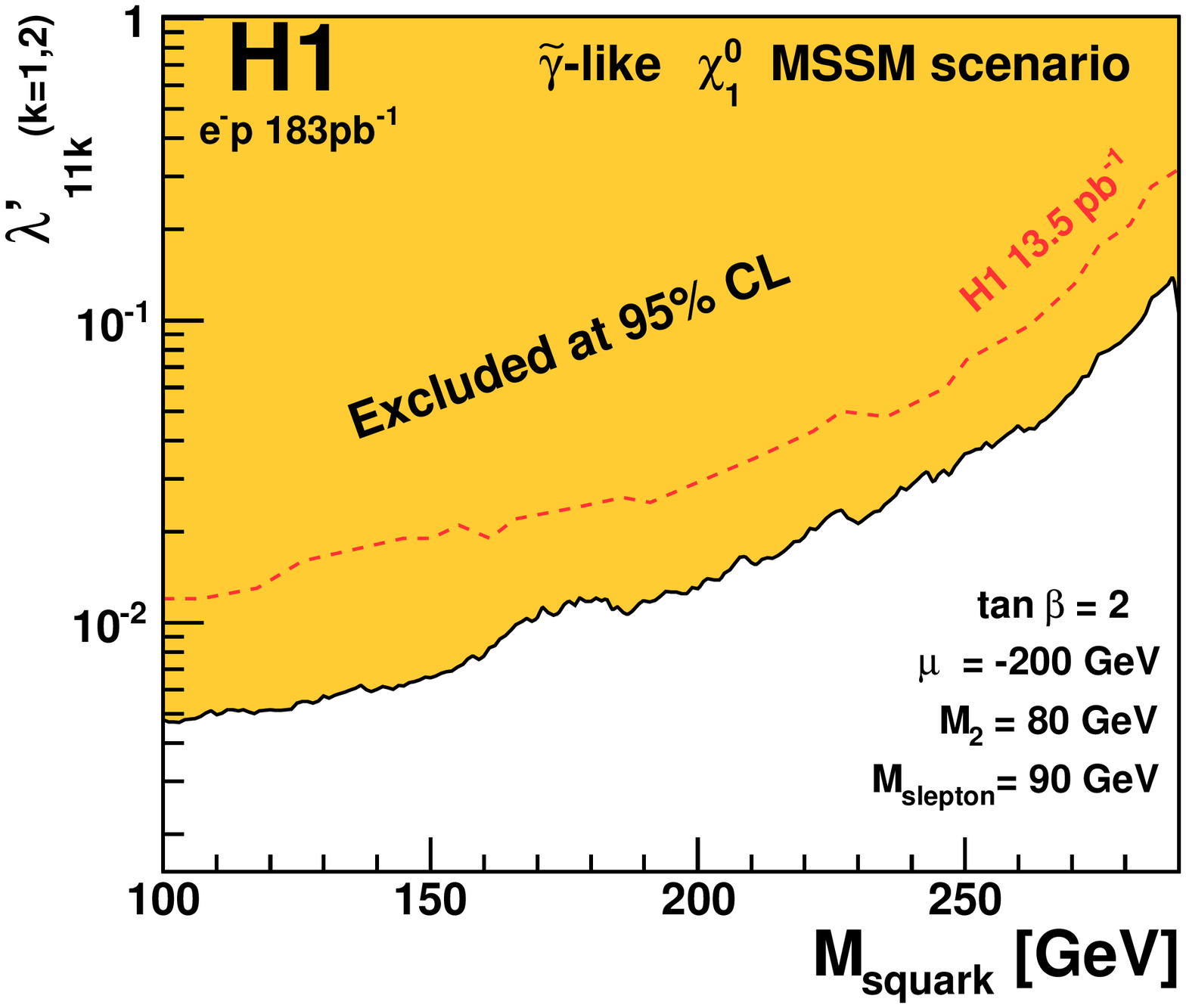,width=0.45\textwidth}
    \epsfig{file=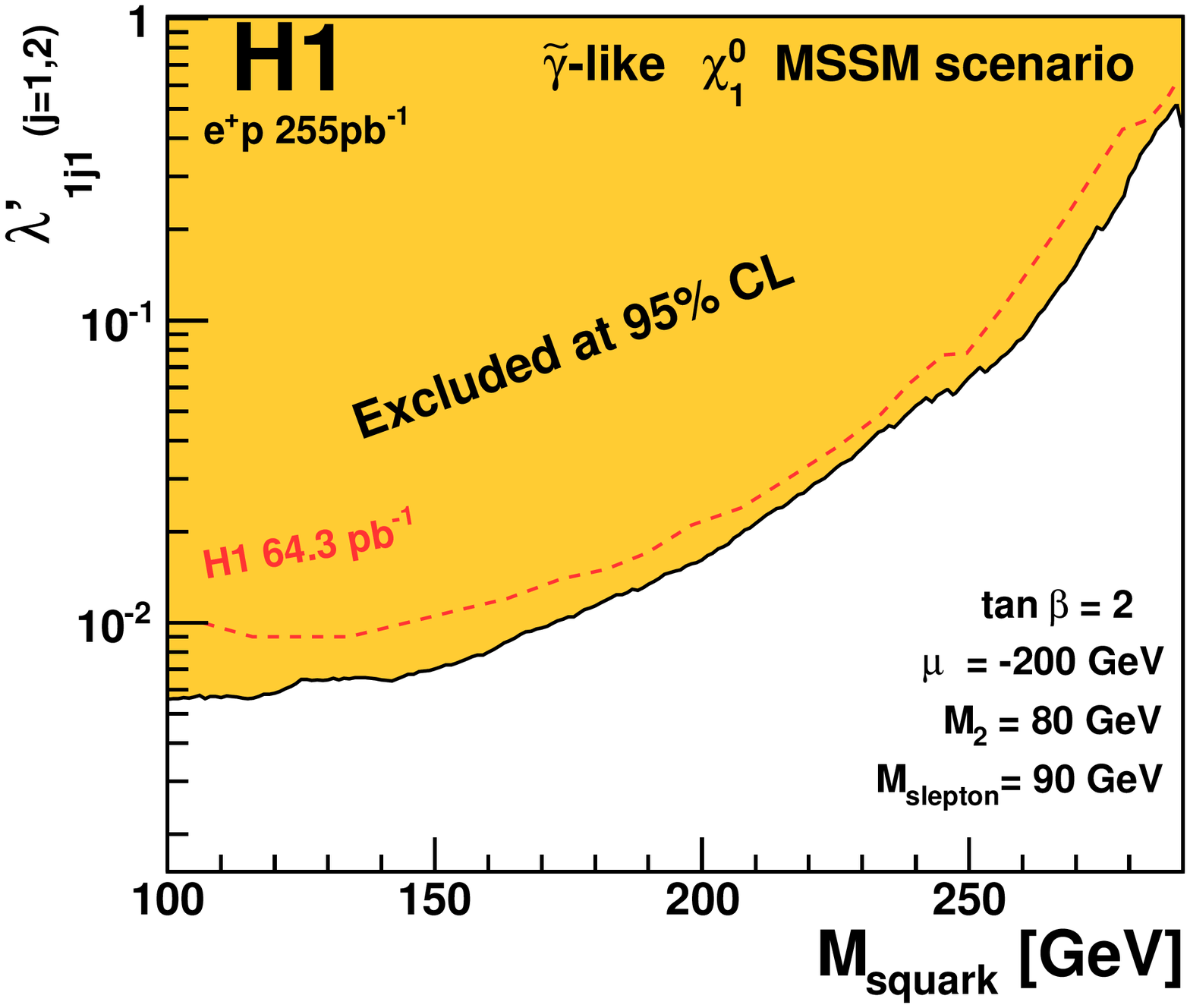,width=0.45\textwidth}\\
    \vspace{1cm}
    \epsfig{file=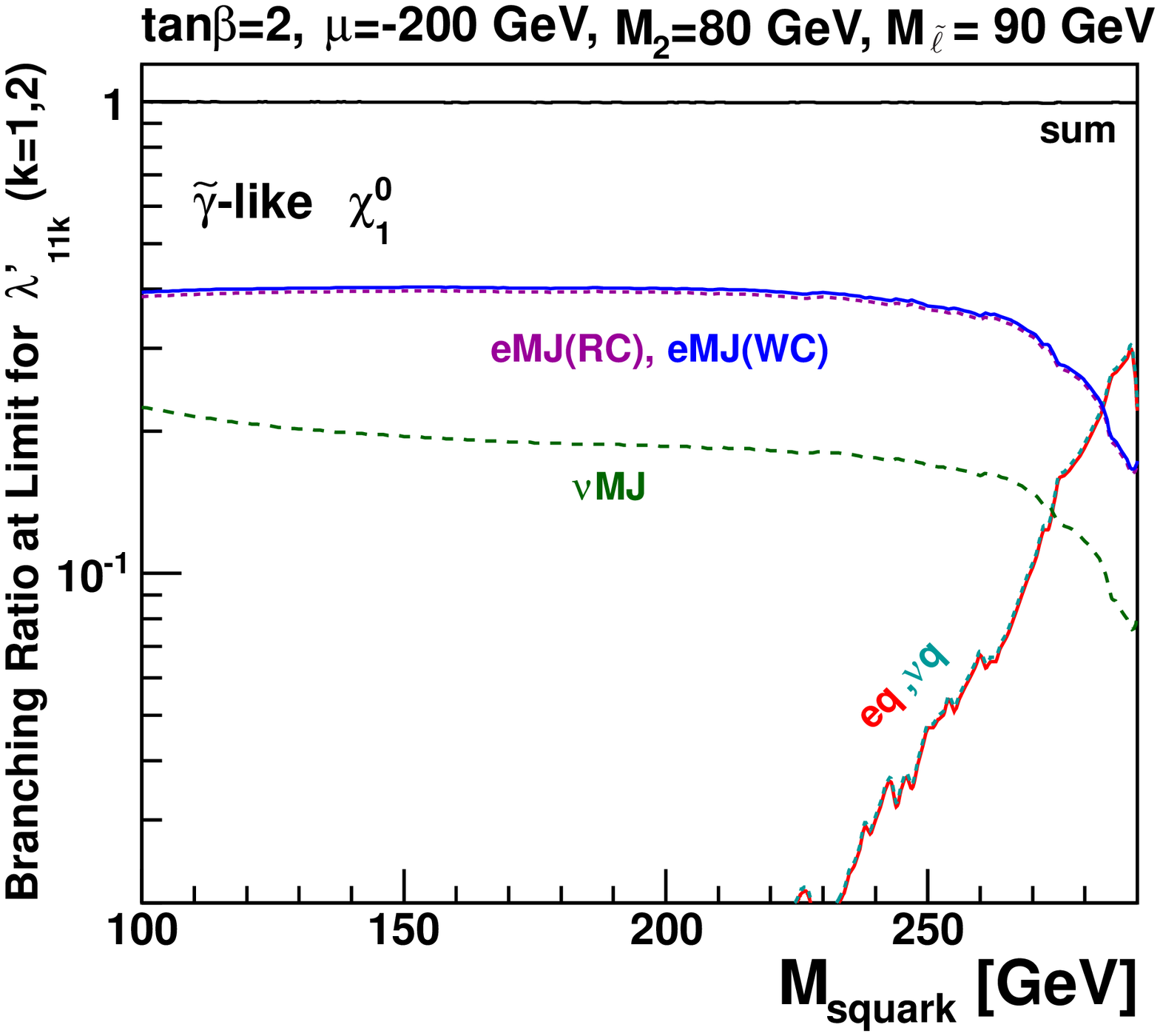,width=0.45\textwidth}
    \epsfig{file=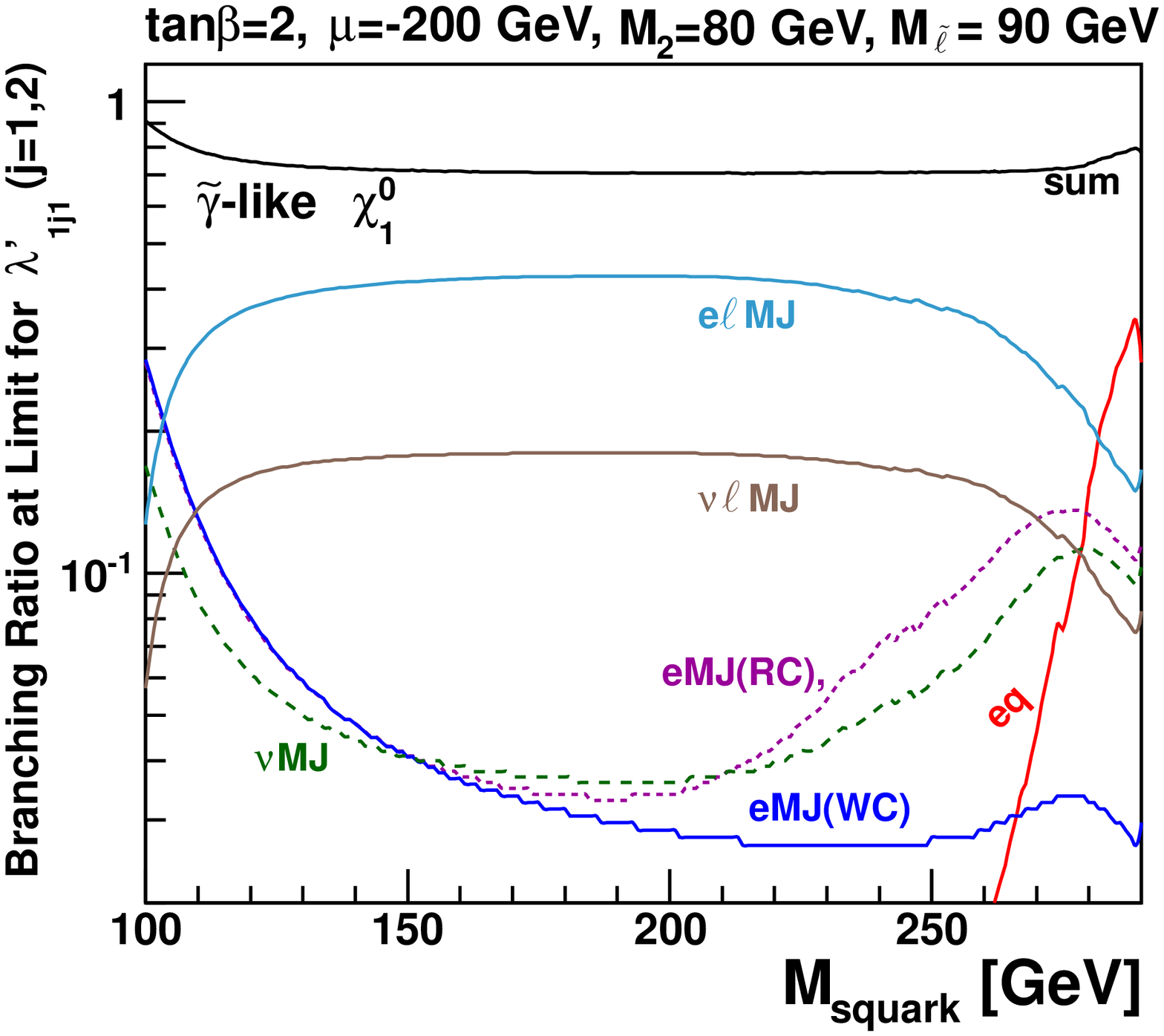,width=0.45\textwidth}

  \end{center}
     \begin{picture}(1,2)(1,2)
\put(21,94){(a)}
\put(95,94){(b)}
\put(21,23){(c)}
\put(95,23){(d)}
\put(27,143){\textsf{\textbf{MSSM scenario with photino-dominated lightest neutralino}}} 
\end{picture}
  \caption{Exclusion limits at $95\%$~CL on (a) $\lambda'_{11k}\,(k=1,2)$ and 
  on (b) $\lambda'_{1j1}\,(j=1,2)$ in 
  a phenomenological MSSM with a photino ($\tilde{\gamma}$) like 
  neutralino ($\chi^0_1$).  For comparison, the corresponding
limit from the previous H1 analysis~\cite{Aktas:2004ij} is also indicated. 
  Also shown are branching ratios to the decay channels considered in this
  analysis for (c) ${\lambda}'_{11k}$ and  (d) ${\lambda}'_{1j1}$ values at the observed limit. 
}
  \label{fig:photinolikeelec}

\end{figure}

For a single point in the parameter space, characterised by $\mu=-200$~GeV, $M_2=80$~GeV and $\tan\beta=2$,
 constraints on the strength of the \Rp\ couplings 
depending on the mass of the squark are derived for
 $\tilde{d}^k_R\,(k=1,2)$ (figure~\ref{fig:photinolikeelec}a) and $\tilde{u}^j_L\,(j=1,2)$
(figure~\ref{fig:photinolikeelec}b) production. 
The HERA sensitivity allows tests of $\lambda'$ values
as low as $10^{-2}$ for squark masses of $100$~GeV. 
For high squark masses the 
sensitivity degrades since the production cross section decreases strongly. 
The limits
from the previous H1 analysis on a smaller data sample~\cite{Aktas:2004ij}
are also indicated.

This choice of parameters
 leads to a dominant photino~($\tilde{\gamma}$) component
to the neutralino's composition. As a consequence, gauge decays are
 likely to result in charged leptons in the final state.
The branching ratios into the decay topologies are shown at the observed 
limit for $\tilde{d}^k_R\,(k=1,2)$ (figure~\ref{fig:photinolikeelec}c) and $\tilde{u}^j_L\,(j=1,2)$
(figure~\ref{fig:photinolikeelec}d) production. 
For $\tilde{d}^k_R\,(k=1,2)$ production the channels $eM\!J$ (RC) and $eM\!J$ (WC) 
each contribute about $40\%$ over a wide range of squark masses and only $10\%$ of 
squark decays appear in the $\nu M\!J$ channel. For squark masses 
approaching the kinematic limit of
the centre--of--mass energy
 the lepton--quark channels $eq$ and $\nu q$
begin to dominate the decays of the squarks, because gauge decay modes become negligible
at high values of the \Rp\ couplings. 
Over the whole mass range the sum of analysed branching ratios is close to $100\%$.
For $\tilde{u}^j_L\,(j=1,2)$ production the channels $e \ell M\!J$ have the highest branching ratio
 over a wide mass range at the observed limit. 

\par
\begin{figure}[t] 
  \begin{center}
    \epsfig{file=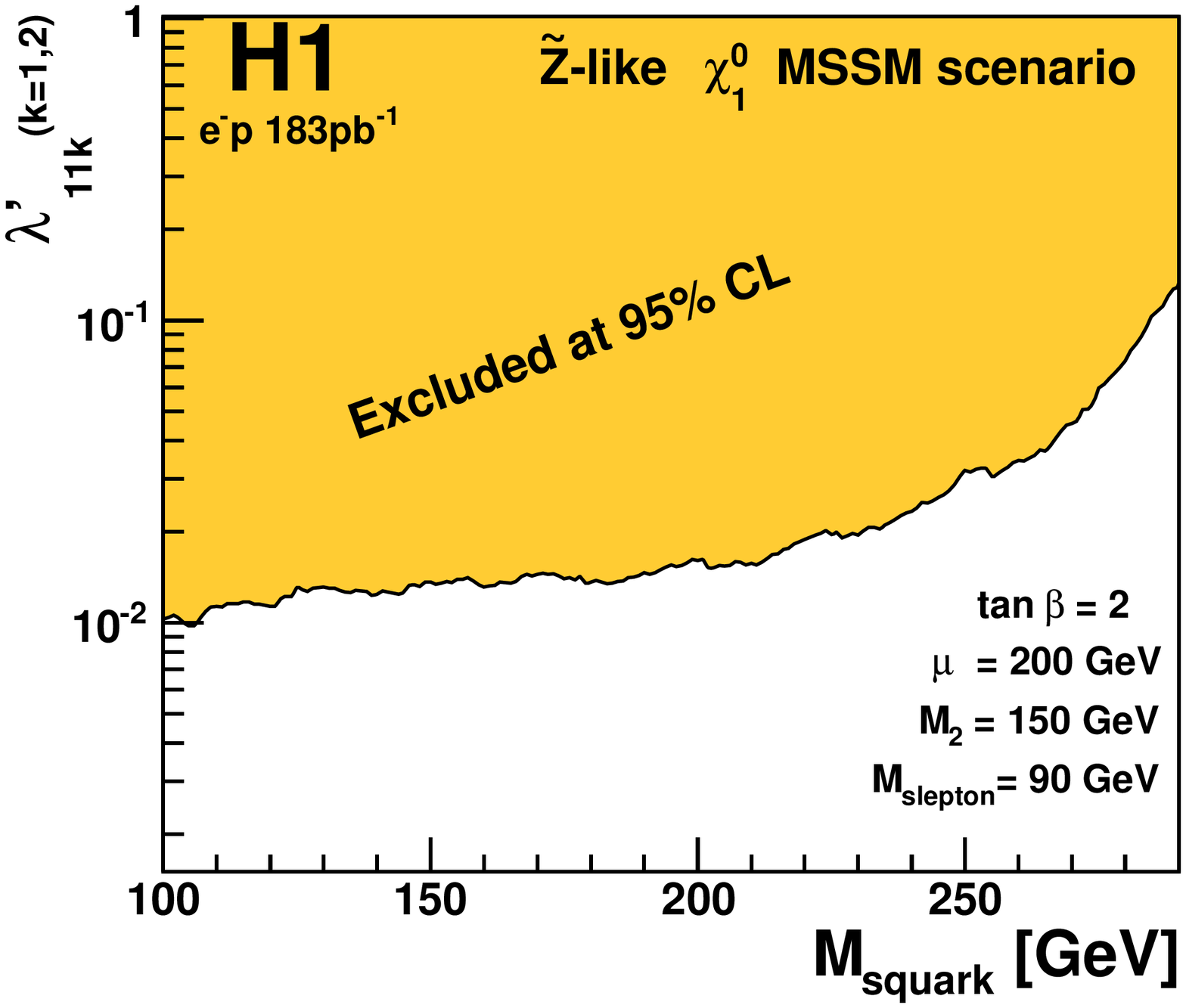,width=0.45\textwidth}
    \epsfig{file=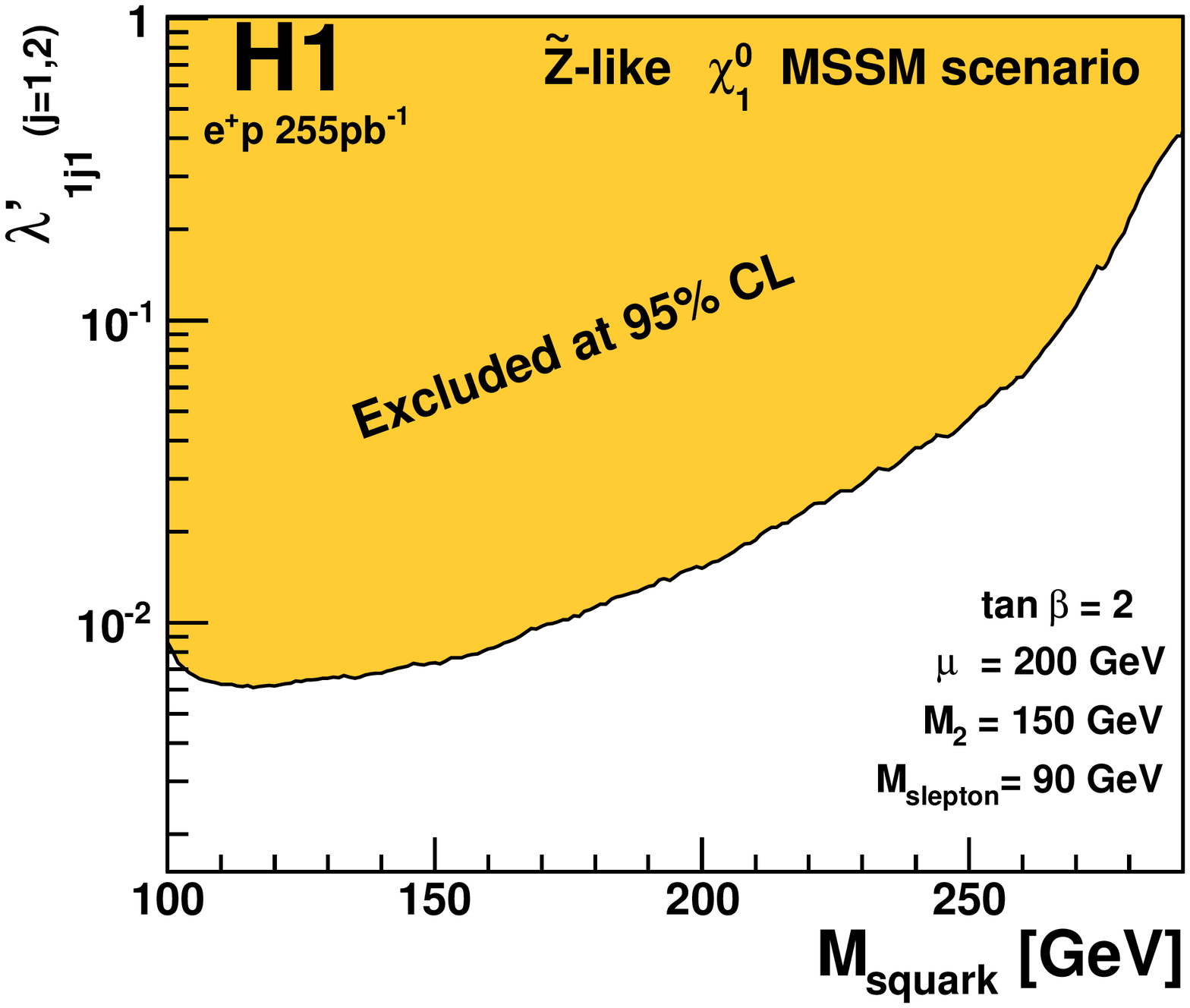,width=0.45\textwidth}\\
    \vspace{1cm}
    \epsfig{file=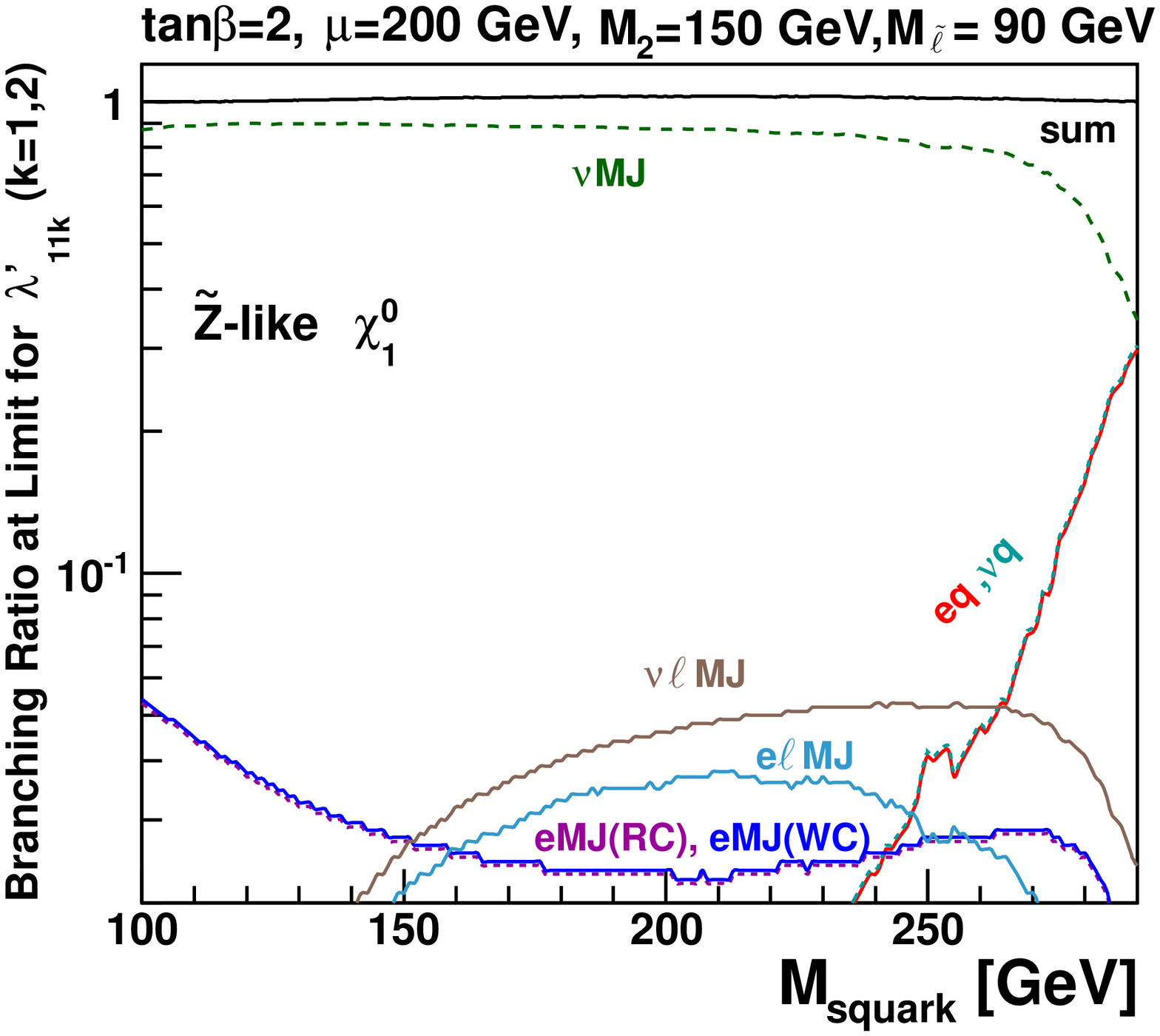,width=0.45\textwidth}
    \epsfig{file=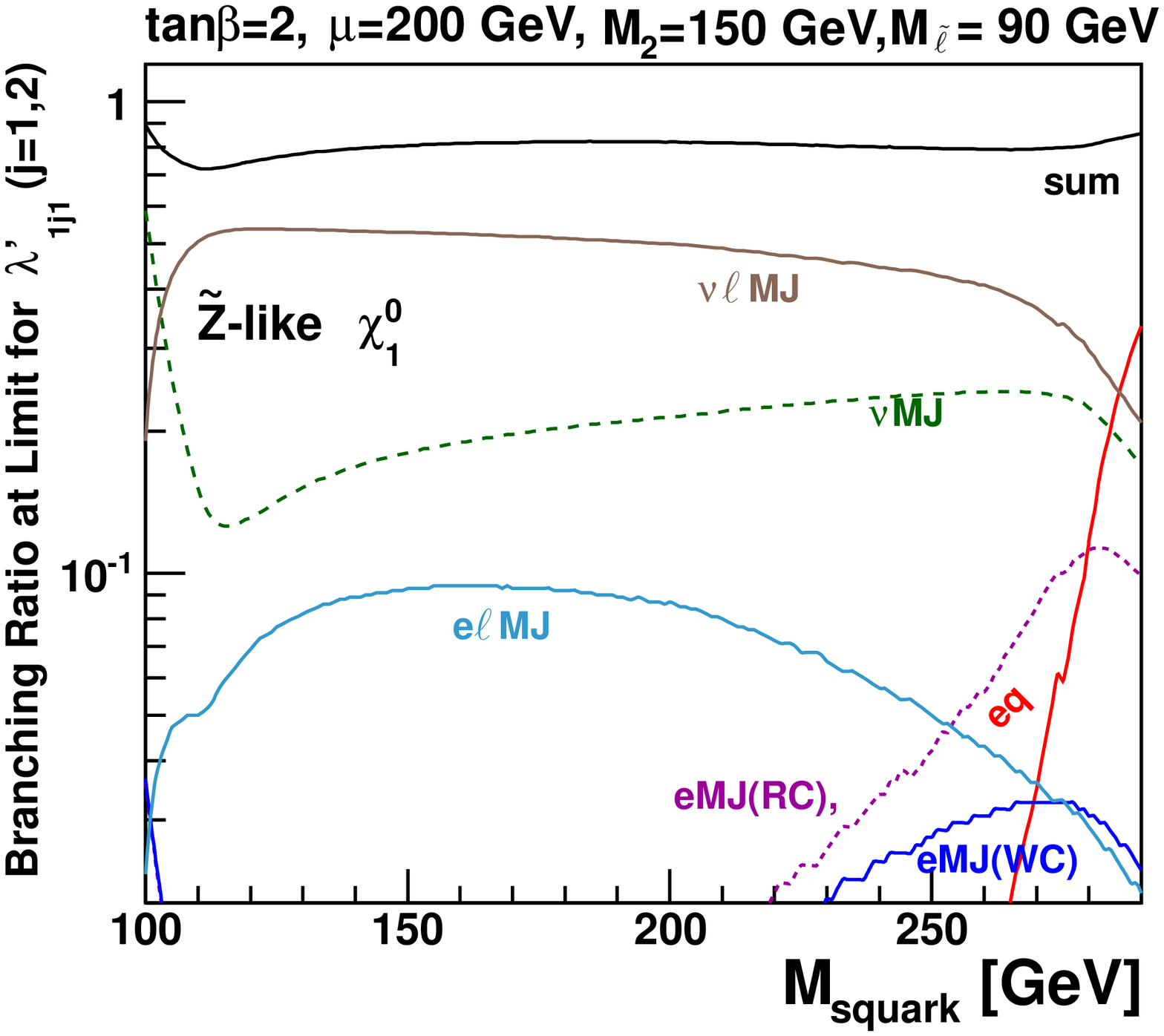,width=0.45\textwidth}

  \end{center}
    \begin{picture}(1,2)(1,2)
\put(21,94){(a)}
\put(95,94){(b)}
\put(21,23){(c)}
\put(95,23){(d)}
\put(30,143){\textsf{\textbf{MSSM scenario with zino-dominated lightest neutralino}}} 

\end{picture}

  \caption{Exclusion limits at $95\%$~CL on (a) $\lambda'_{11k}\,(k=1,2)$ and on 
  (b) $\lambda'_{1j1}\,(j=1,2)$ in 
  a phenomenological MSSM with a zino ($\tilde{Z}$) like neutralino
  ($\chi^0_1$). Also shown are branching ratios to the decay channels considered in this
  analysis for (c) ${\lambda}'_{11k}$ and  (d) ${\lambda}'_{1j1}$ values at the observed limit. 
}
  \label{fig:zinolikeelec}

\end{figure}

A different point in the parameter space 
yields a complementary scenario with the choice of $\mu=200$~GeV,
$M_2=150$~GeV and $\tan\beta=2$.
Again the neutralino $\chi^0_1$ is the LSP but
its composition is now dominated by a zino ($\tilde{Z}$) component.
Squark decays are now more likely to produce neutrinos in the final state.  
The constraints on the couplings depending on the mass of the squark 
are shown 
for $\tilde{d}^k_R\,(k=1,2)$ (figure~\ref{fig:zinolikeelec}a) and $\tilde{u}^j_L\,(j=1,2)$
(figure~\ref{fig:zinolikeelec}b) production and are of  
the same order of magnitude as in the photino scenario.
Branching ratios at the
observed limits show dominant contributions 
from the $\nu M\!J$ and $\nu \ell M\!J$
channels~(figures~\ref{fig:zinolikeelec}c and~\ref{fig:zinolikeelec}d). 
\par
\begin{figure}[t] 
  \begin{center}
    \epsfig{file=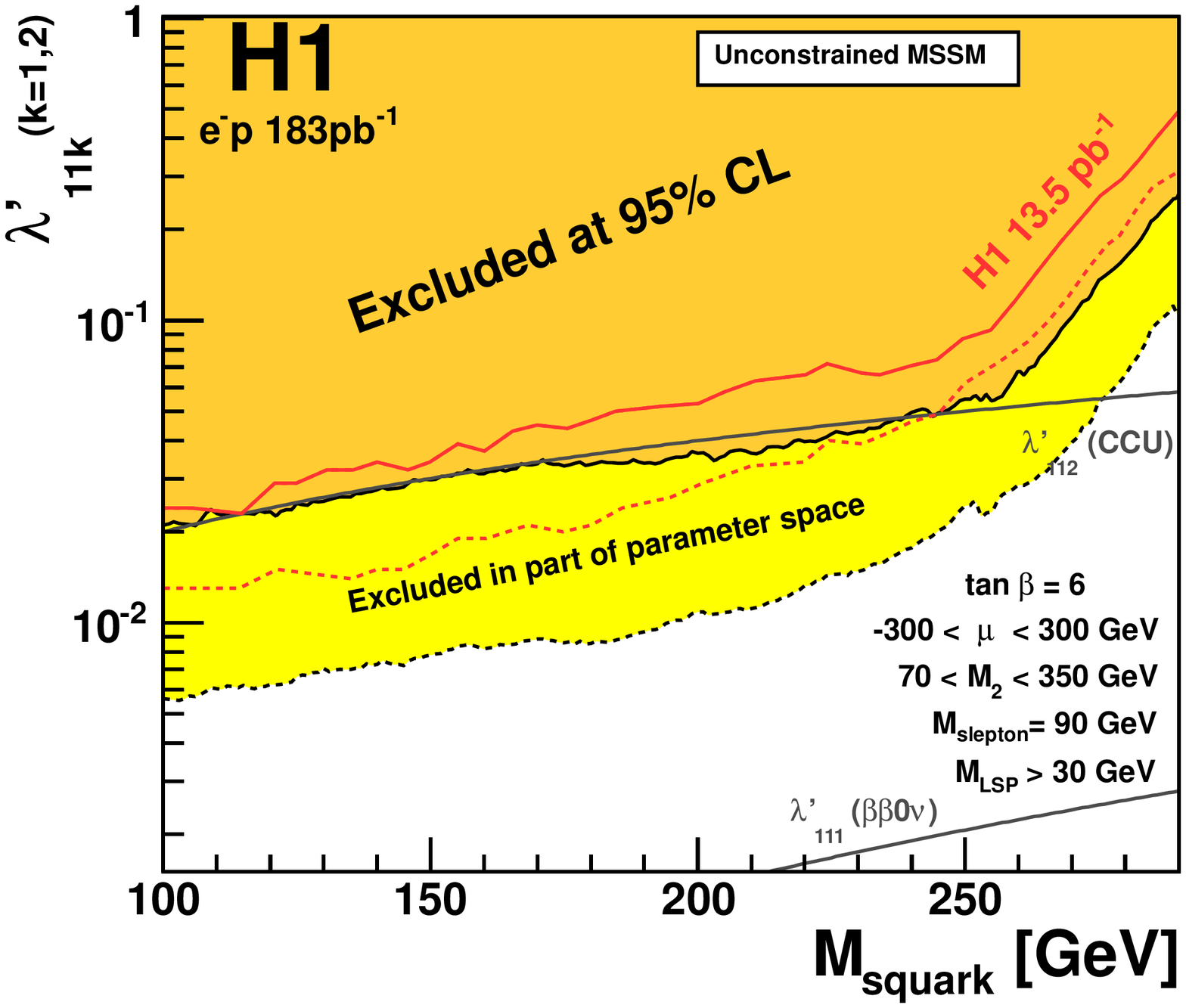,width=0.65\textwidth}\\
    \epsfig{file=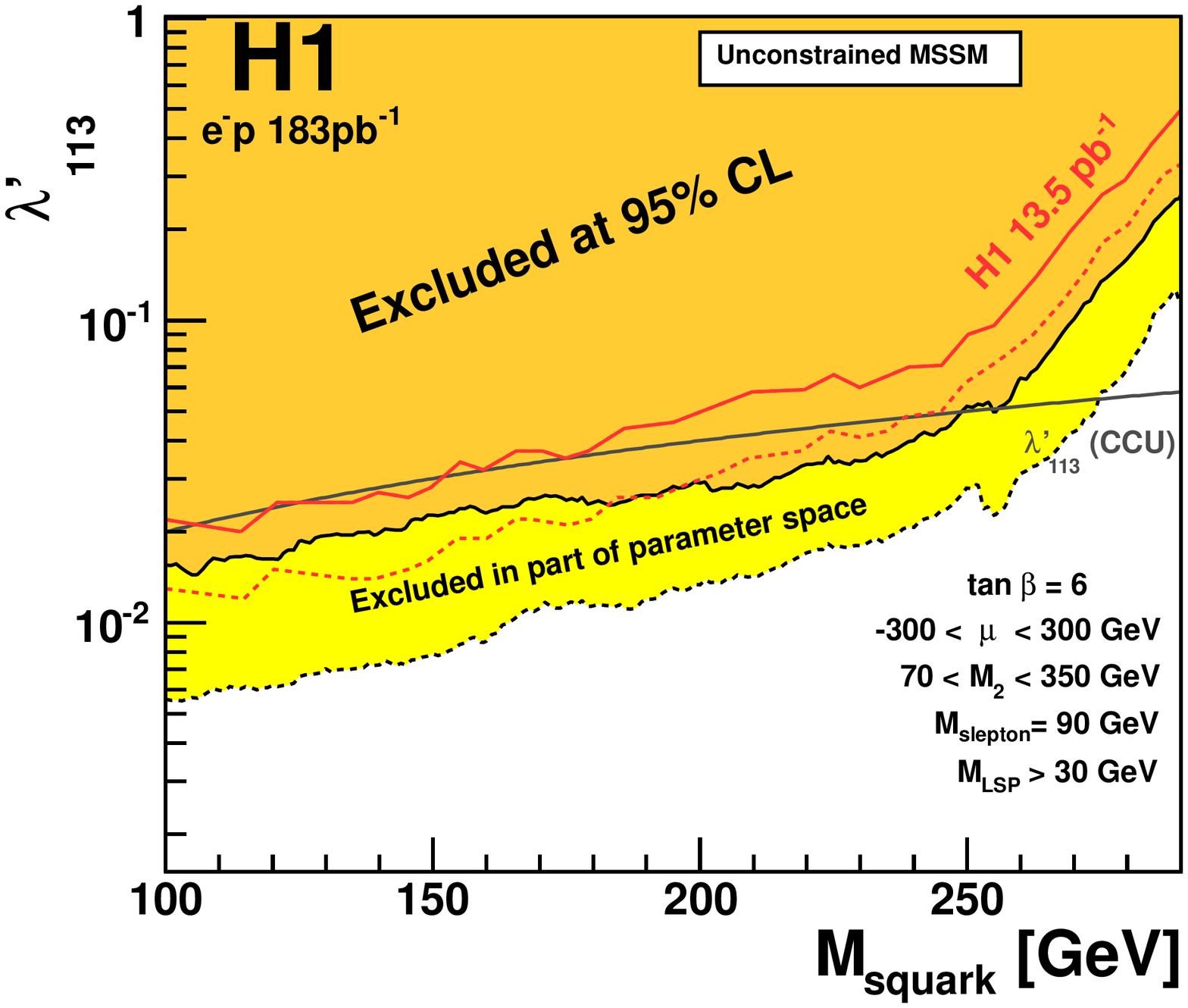,width=0.65\textwidth}
  \end{center}
  \begin{picture}(1,2)(1,2)
\put(47,116){(a)}
\put(47,28){(b)}
\end{picture}

  \caption{Exclusion limits ($95\,\%$ CL) on $\lambda'_{11k}$ for (a) $k=1,2$ 
and (b) $k=3$ 
as a function of the squark mass from a scan of the MSSM parameter space.
The dark filled region indicates values of the coupling $\lambda'_{11k}$ excluded in all 
investigated scenarios whereas the light filled region is excluded only in part of the scenarios. 
Indirect limits from 
neutrinoless double beta decay experiments ($\beta\beta 0\nu$)~\cite{BETA0NU,barbier} and 
tests of charged current 
universality (CCU)~\cite{CCU,barbier} are also shown. For comparison, the corresponding
limits from the previous H1 analysis~\cite{Aktas:2004ij} are also indicated.}
  \label{fig:scan11k}
\end{figure}
\begin{figure}[t] 
  \begin{center}
    \epsfig{file=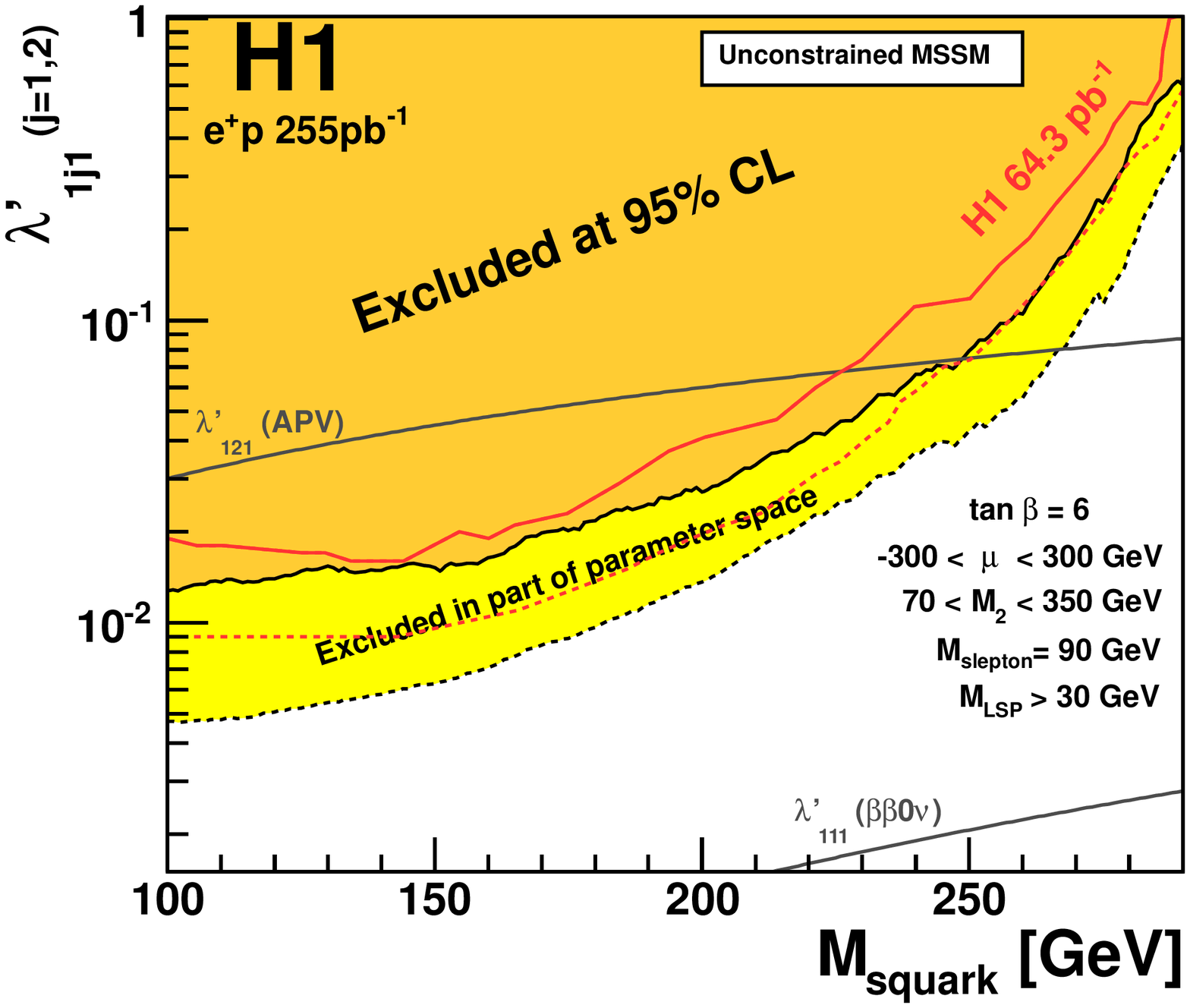,width=0.65\textwidth}\\
    \epsfig{file=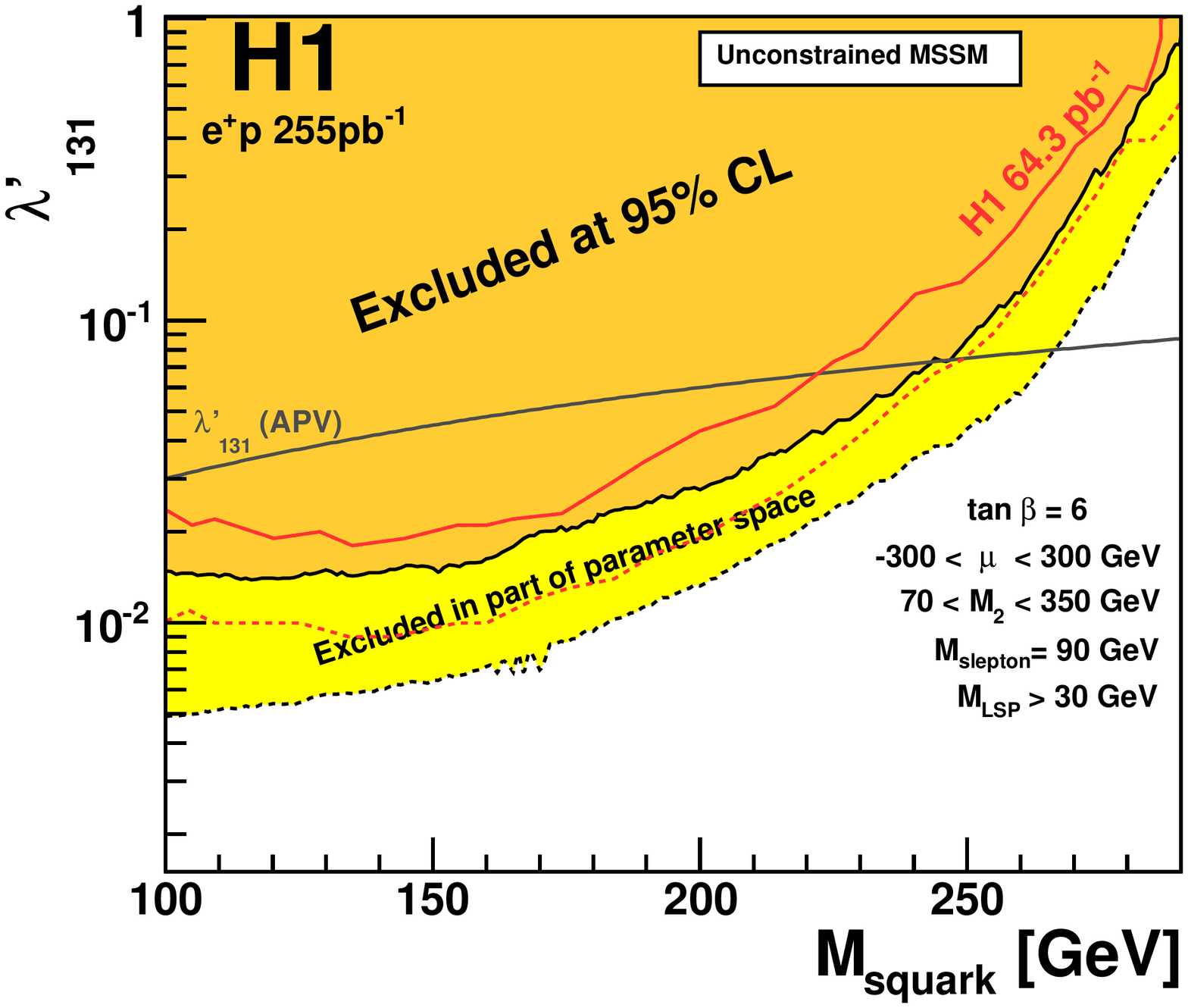,width=0.65\textwidth}
  \end{center}
\begin{picture}(1,2)(1,2)
\put(47,116){(a)}
\put(47,28){(b)}
\end{picture}

  \caption{Exclusion limits  ($95\,\%$ CL) on $\lambda'_{1j1}$ for (a) $j=1,2$ 
and (b) $j=3$ 
as a function of the squark mass from a scan of the MSSM parameter space as 
indicated in the figures. 
The dark filled region indicates values of the coupling $\lambda'_{1j1}$ excluded in all 
investigated scenarios whereas the light filled region is excluded only in part of the scenarios. 
Indirect limits from neutrinoless 
double beta decay experiments ($\beta\beta 0\nu$)~\cite{BETA0NU,barbier} and atomic parity violation 
(APV)~\cite{APV,barbier} 
are also shown.
For comparison, the corresponding
limits from the previous H1 analysis~\cite{Aktas:2004ij} are also indicated.}
  \label{fig:scan1j1}
\end{figure}

These two scenarios illustrate the sensitivity for various model configurations 
achieved by
the combination of the complementary search topologies. 
The sensitivity of the analysis is explored in a scan of the MSSM parameters.
The parameters $M_2$ and $\mu$  are varied in the range 
 $70$~GeV~$< M_2 < 350$~GeV 
 and $-300$~GeV~$ < \mu < 300$~GeV  for $\tan \beta = 6$. 
 Parameter sets leading to a scalar LSP or to 
 LSP masses below $30$~GeV are not considered. The latter restriction, as well 
 as the lower 
 boundary of the $M_2$ range, are motivated by the exclusion domains resulting from gaugino searches in 
 \Rp\ SUSY at LEP~\cite{L3RPV}. 
Figures~\ref{fig:scan11k} and \ref{fig:scan1j1} show the resulting constraints on the couplings as a function of the squark mass. The region of values excluded at $95\%$~CL for the couplings in all scenarios and the best exclusion limit achieved in all scenarios are indicated 
for first and second generation squarks $\tilde{d}_R$, $\tilde{s}_R$ (figure~\ref{fig:scan11k}a) and
$\tilde{u}_L$, $\tilde{c}_L$ (figure~\ref{fig:scan1j1}a)
as well as for third generation squarks $\tilde{b}_R$ (figure~\ref{fig:scan11k}b)
and $\tilde{t}_L$ (figure~\ref{fig:scan1j1}b). 
The resulting exclusion domains are compared to the previous H1 
results~\cite{Aktas:2004ij}.
Constraints on the \Rp\ couplings are also available as indirect limits from 
low energy experiments probing virtual squark contributions~\cite{barbier}.
The production of up--type and down--type squarks via the $\lambda'_{111}$ coupling 
is strongly constrained
by the non--observation of neutrinoless double beta decay ($\beta \beta 0 \nu$)~\cite{BETA0NU,barbier}.
The best indirect limit on the couplings $\lambda'_{112}$ and $\lambda'_{113}$ results from
tests of charged current universality (CCU)~\cite{CCU,barbier} and can be compared to the 
direct limits obtained in this analysis for $\lambda'_{11k}$ in figure~\ref{fig:scan11k}.
The best indirect limit on the couplings $\lambda'_{121}$ and $\lambda'_{131}$ comes from
atomic parity violation (APV) measurements~\cite{APV,barbier} and can be compared to the 
direct limits obtained for $\lambda'_{1j1}$  in figure~\ref{fig:scan1j1}. 

In the part of the parameter space considered here, Yukawa
couplings of electromagnetic strength $\lambda'_{1j1}$ or $\lambda'_{11k}=\sqrt{4\pi\alpha_{\rm em}}= 0.3$,
are excluded up to masses of $275$~GeV
 at $95\%$~CL for up--type squarks and up to
masses of $290$~GeV for down--type squarks.

\subsection{Constraints on the Minimal Supergravity Model}
Constraints are also obtained on the Minimal Supergravity Model (mSUGRA)~\cite{MSUGRA}
which is a complete SUSY model using the assumption of gauge coupling unification
and radiative electroweak symmetry breaking (REWSB) with the choice of 5 parameters: the common mass of scalar sparticles $m_0$;
the common mass of fermionic sparticles $m_{1/2}$; the common trilinear coupling $A_0$;
the ratio of Higgs vacuum expectation values $\tan \beta$; and the sign of the Higgs mixing parameter $\mu$.
The masses of squarks, sleptons and gauginos as well as the branching ratios in the analysis channels are determined by the set
 ($m_0,m_{1/2},\tan\beta,sign(\mu),A_0$) for given values of the couplings $\lambda'_{11k}$ and $\lambda'_{1j1}$.
The program SUSPECT~2.1~\cite{SUSPECT} is used to obtain the REWSB solution for 
$|\mu|$ and to calculate the full supersymmetric mass spectrum.
$A_0$ enters only marginally in the interpretation and is set to zero. 
The parameter $\mu$ is taken with negative sign.

\begin{figure}[t] 
  \begin{center}
    \epsfig{file=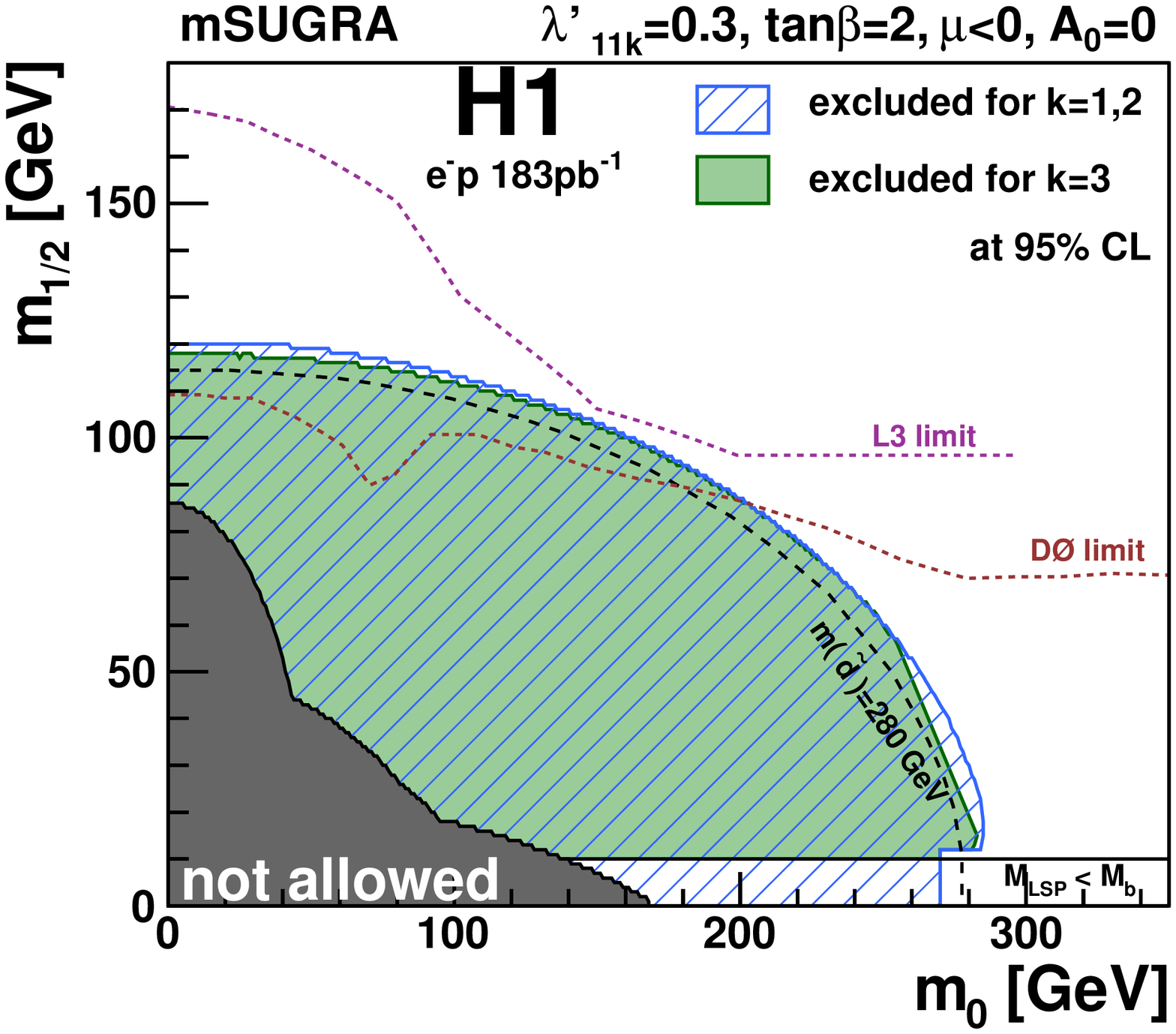,width=0.65\textwidth}\\
    \epsfig{file=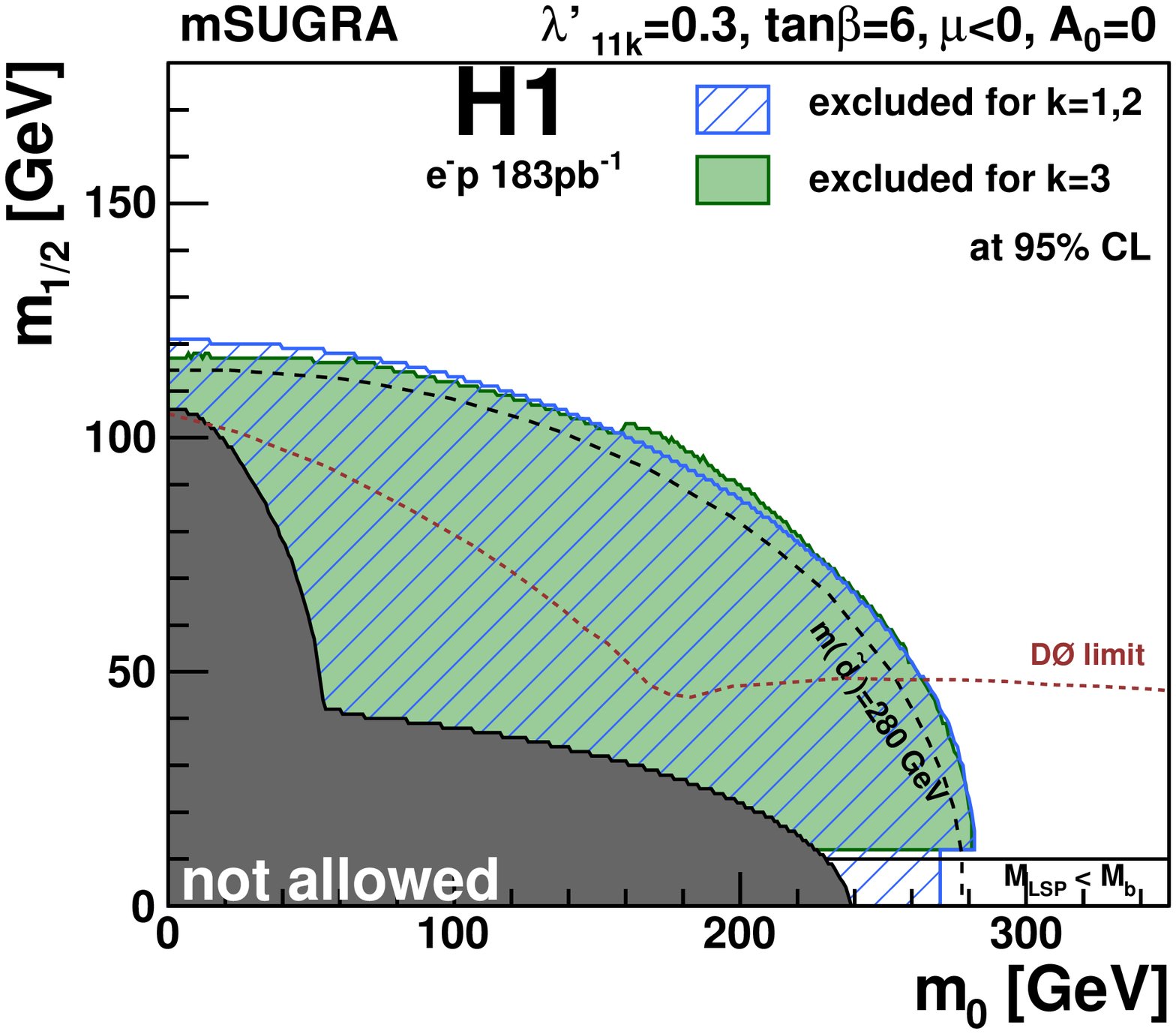,width=0.65\textwidth}
  \end{center}
  \begin{picture}(1,2)(1,2)
\put(60,177){(a)}
\put(60,90){(b)}
\end{picture}

  \caption{Exclusion limits  ($95\,\%$ CL) in the $m_0,m_{1/2}$ plane
assuming $\lambda'_{11k}=0.3$ for (a) $\tan \beta=2$ and (b) $\tan\beta=6$
  for $k=1,2$ (hatched region) and $k=3$ (light filled region). A curve of constant squark
  mass is illustrated for $m(\tilde{d})=280$~GeV.
Also indicated are constraints obtained by the L3 experiment at LEP~\cite{L3RPV} 
and the D\O\ experiment at the Tevatron~\cite{D0RES}. The dark filled region labelled 
as ``not allowed'' indicates 
where no REWSB solution is possible or where the LSP is a sfermion.
}
  \label{fig:msugra11k}
\end{figure}

\begin{figure}[t] 
  \begin{center}
    \epsfig{file=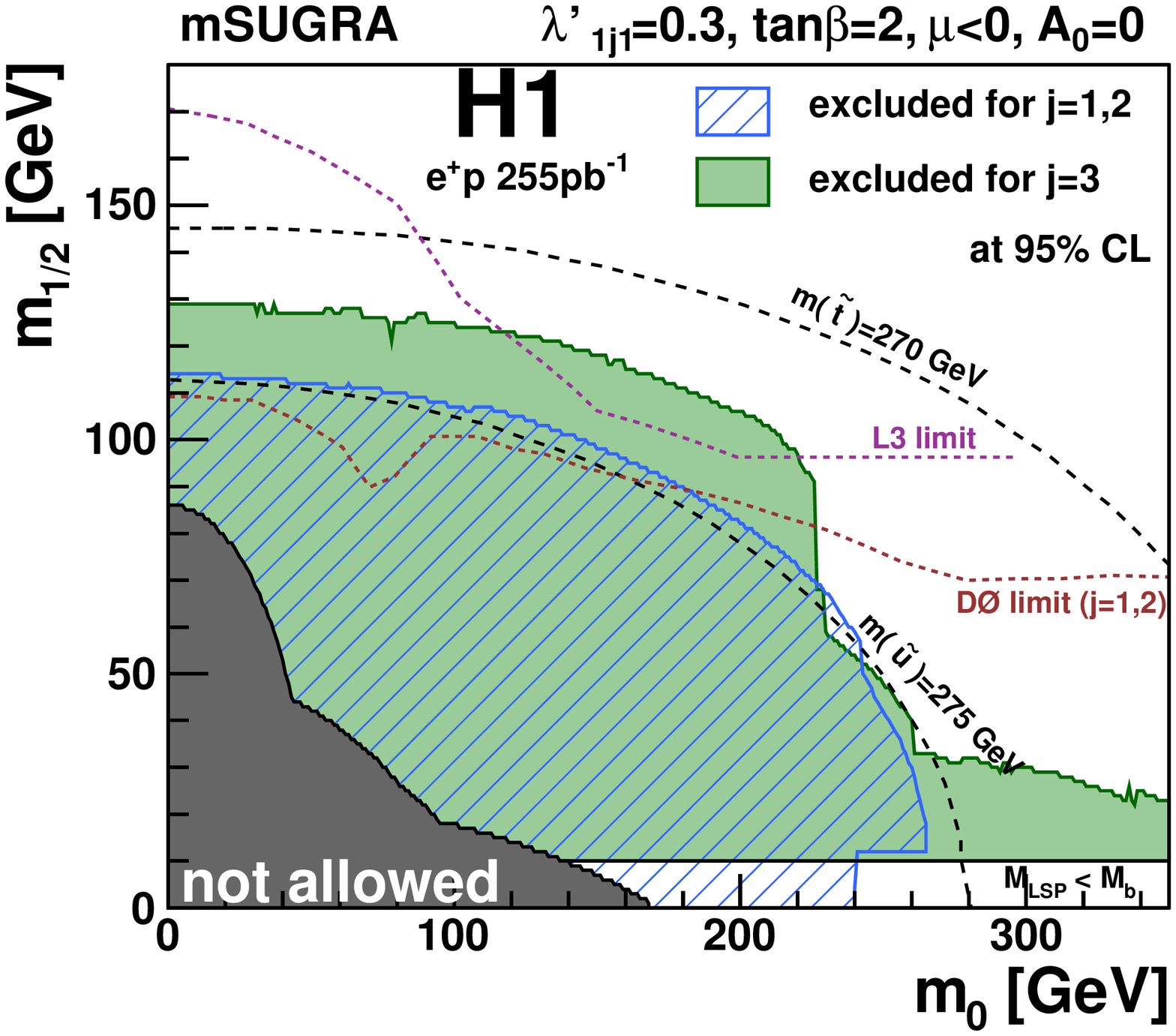,width=0.65\textwidth}\\
    \epsfig{file=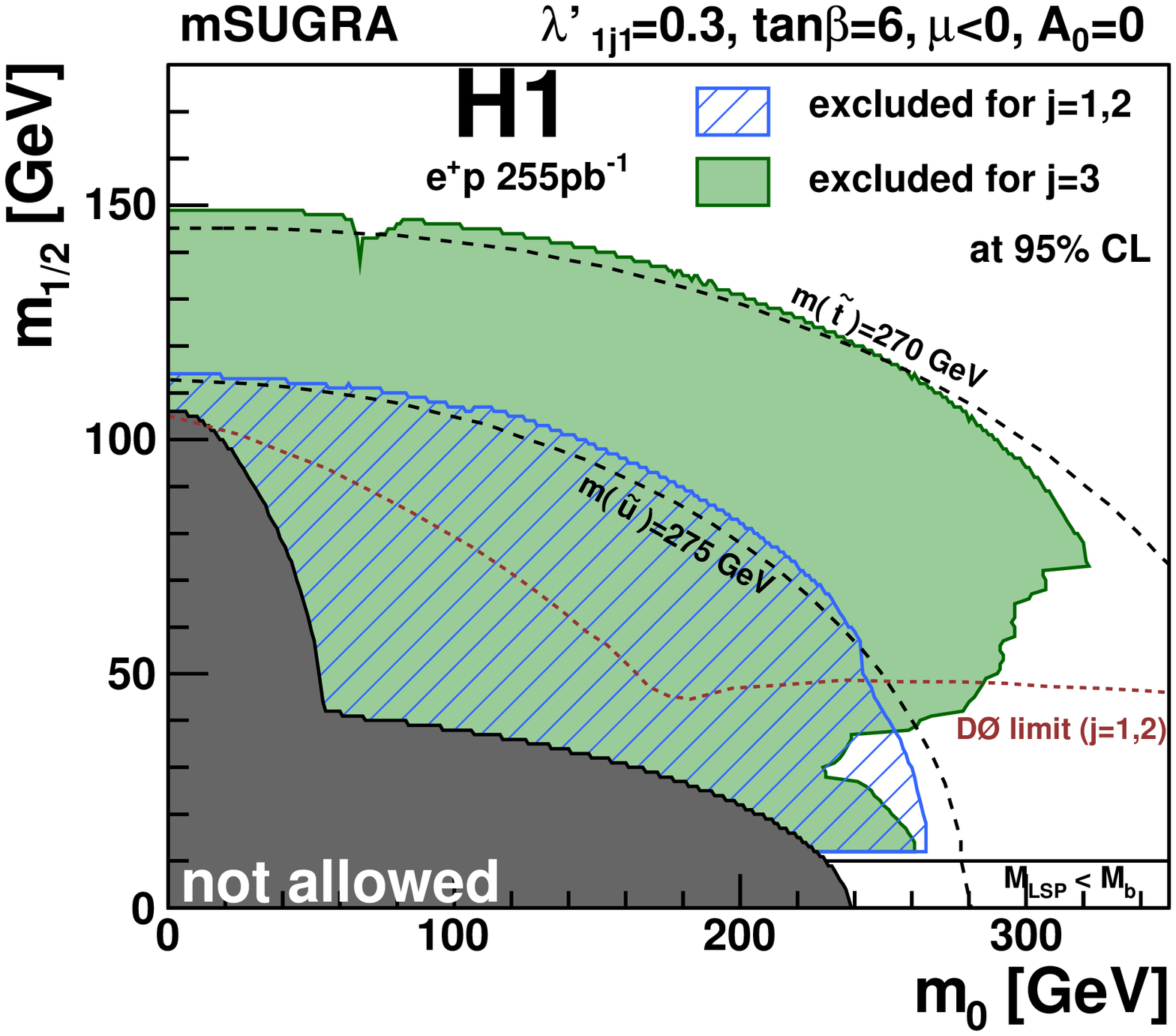,width=0.65\textwidth}
  \end{center}
  \begin{picture}(1,2)(1,2)
\put(60,177){(a)}
\put(60,90){(b)}
\end{picture}

  \caption{Exclusion limits  ($95\,\%$ CL) in the $m_0,m_{1/2}$ plane
assuming $\lambda'_{1j1}=0.3$ for (a) $\tan \beta=2$ and (b) $\tan\beta=6$
  for $j=1,2$ (hatched region) and $j=3$ (light filled region). Curves of constant squark
  mass are illustrated for $m(\tilde{u})=275$~GeV and $m(\tilde{t})=270$~GeV.
Also indicated are constraints obtained by the L3 experiment at 
LEP~\cite{L3RPV} and the D\O\ experiment at the Tevatron~\cite{D0RES}. The 
dark filled region labelled as ``not allowed'' indicates 
where no REWSB solution is possible or where the LSP is a sfermion.
}

  \label{fig:msugra1j1}
\end{figure}

Figures~\ref{fig:msugra11k} and~\ref{fig:msugra1j1} show constraints in the \mbox{$m_0,m_{1/2}$} plane 
when values of the couplings are assumed to be of the electromagnetic coupling strength 
$\lambda'_{11k}=0.3$ or $\lambda'_{1j1}=0.3$
for different values of $\tan \beta$. 
The excluded region typically covers masses of $m(\tilde{u})=275$~GeV,  $m(\tilde{t})=270$~GeV
and $m(\tilde{d})=280$~GeV, as indicated in the figures.
Complementary constraints are obtained by the L3 experiment~\cite{L3RPV} at LEP and the 
\mbox{D\O} experiment~\cite{D0RES} at the 
Tevatron which exploit di--electron events.
The LEP and Tevatron limits are independent of the Yukawa coupling.
For $\tan \beta =2$, the parameter space is more 
strongly constrained by the searches for 
gauginos and sleptons at the L3 experiment at LEP, as shown in 
figures~\ref{fig:msugra11k} and \ref{fig:msugra1j1}. This is the only $\tan \beta$ value considered 
in the L3 analysis, although results for higher values are expected to be similar~\cite{L3RPV}.
Compared to the \mbox{D\O} experiment, the H1 
limits are more stringent only for low values of $m_0$ for $\tan \beta =2$ , whereas for $\tan \beta = 6$ 
the domain excluded by H1
is considerably larger.

\begin{figure}[t] 
  \begin{center}
    \epsfig{file=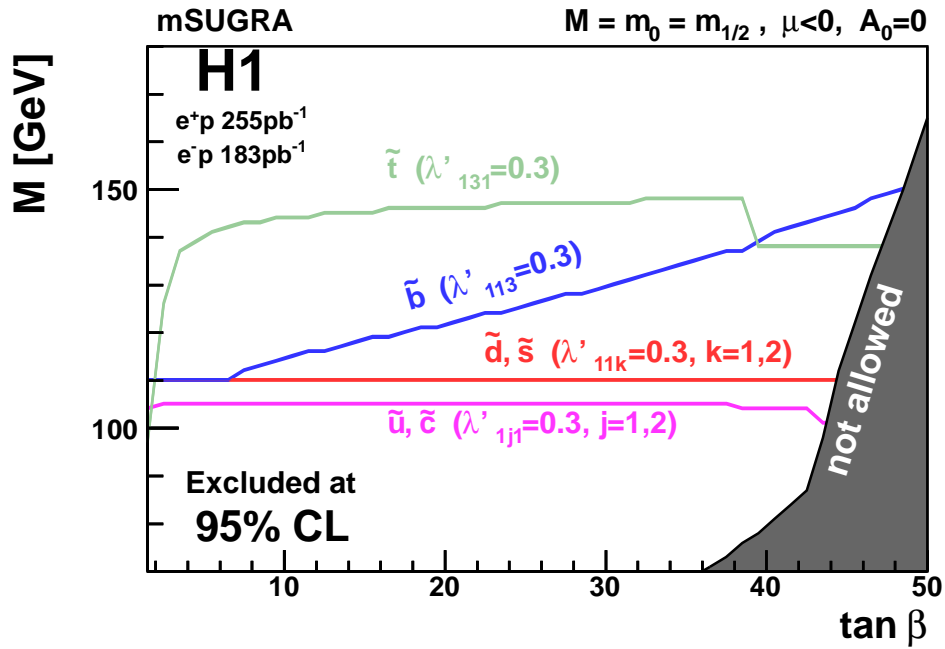 ,width=0.8\textwidth}
  \end{center}
  \caption{
Exclusion limits for $M=m_0=m_{1/2}$ in mSUGRA as function of $\tan \beta$. Shown
are the $95\,\%$ CL exclusion domains for the model parameters
 from the production of squarks of first and second
generation ($\tilde{u},\tilde{c}$ and $\tilde{d},\tilde{s}$) and of third generation ($\tilde{t},\tilde{b}$)
 assuming a value of $\lambda'=0.3$ for the respective coupling.  The area below the
curves is excluded. The dark filled region region labelled as ``not allowed'' indicates 
where no REWSB solution is possible or where the LSP is a sfermion.
}
  \label{fig:tanbeta}
\end{figure}

The exclusion limits in figure~\ref{fig:msugra11k} are very similar for all 
three flavours of down--type squarks.
Significant differences are observed between the first two and the third
generation of up--type squarks.
 The stronger limit for stop squark production 
 results from
 strong mixing effects that occur for third generation squarks with increasing $\tan \beta$,
 leading to masses for stop squarks lower than for first and second generation up--type squarks.
The $\tan \beta$ dependence of the mSUGRA exclusion limits is studied assuming a unified common mass
$M=m_0=m_{1/2}$ and \Rp\ couplings of electromagnetic coupling strength.
This is illustrated in figure~\ref{fig:tanbeta} for the individual
flavours.
For the first two generations of up--type and down--type squarks no dependence on $\tan \beta$ is observed and
values of $M<105$~GeV and $M<110$~GeV, respectively, are excluded over the whole range.
In the case of stop squark production, significantly higher values (up to $M<148$~GeV) are excluded
 due to the presence of a light stop squark state. This effect
 is also observed for sbottom production, 
 where increased values of $\tan \beta$ allow higher values of $M$  to be excluded.
 While there is a steep increase for the limit from stop squark production at small 
 $\tan \beta$ and a flat
 plateau over the remaining $\tan \beta$ range, the limit from sbottom production 
 increases steadily over the range of $\tan \beta$ values.
 The sharp edge in the stop exclusion curve at $\tan\beta\approx38$ for stop production
follows from
 mixing effects in the $\tilde {\tau}$ sector at high $\tan\beta$ leading to scenarios with
strong contributions from events with $\tau$ leptons in the final state, which are not explicitely considered in
the analysis.

\section{Summary}
A search for $R$--parity violating production of squarks in $255\,\mathrm{pb^{-1}}$ of $e^+ p$ and $183\,\mathrm{pb^{-1}}$ of $e^- p$ 
collisions at HERA is presented. No significant deviation from the Standard Model is 
observed in the study of final state topologies which may result
from direct or indirect \Rp\ squark decays. Mass dependent limits on the \Rp\ couplings 
$\lambda'_{1j1}$ and $\lambda'_{11k}$ ($j,k=1,2,3$) 
are derived within a phenomenological version of the MSSM.
The existence of $\tilde{u}_L$--type and $\tilde{d}_R$--type squarks of all three 
generations with masses up to $275$~GeV and $290$~GeV, respectively,
 is excluded at the $95\%$ CL for Yukawa couplings of electromagnetic strength.
These mass limits set the most stringent direct bounds on 
$\lambda'_{1j1}$ and $\lambda'_{11k}$.
For lower squark masses, the results improve the indirect bounds set by low--energy 
experiments. Exclusion limits are also derived in the mSUGRA model, and are competitive with and complementary to those derived at 
the LEP and Tevatron colliders.

\section*{Acknowledgements}

We are grateful to the HERA machine group whose outstanding
efforts have made this experiment possible. 
We thank the engineers and technicians for their work in constructing and
maintaining the H1 detector, our funding agencies for 
financial support, the
DESY technical staff for continual assistance
and the DESY directorate for support and for the
hospitality which they extend to the non--DESY 
members of the collaboration.

\vfill


\clearpage

\label{sec:limideri}
\begin{table}[t]
  \renewcommand{\arraystretch}{1.2}
 \begin{center}
  \begin{tabular}{|c|cccc|cccc|ccc|}
  \multicolumn{12}{c}{{H1 Search for Squarks in \Rp\ SUSY}} \\
  \hline
  \rule[-2mm]{0mm}{7mm} Selection& \multicolumn{4}{c|}{ $e^-p$ $(183$~pb$^{-1})$ } & \multicolumn{4}{c|}{  $e^+p$ $(255$~pb$^{-1})$ } & \multicolumn{3}{c|}{Range of Signal} \\  
\multicolumn{1}{|c|}{ Channel} & \multicolumn{1}{c}{ Data} &  \multicolumn{3}{c|}{SM Expectation} & { Data} &  \multicolumn{3}{c|}{ SM Expectation} &  \multicolumn{3}{c|}{ Efficiencies}\\ 
\hline \hline
$eq$ 	 
	 & $   3121     $ & $ 3215 $ & $\pm $ & $   336 $ 
	 	& $   2946     $ & $ 2899 $ & $\pm $ & $  302 $  
	 & $30\%$&$-$&$40\%$\\
$\nu q$ 	 
	 & $   2858     $ & $ 2983 $ & $\pm $ & $   358 $   
	 	&     --     & & -- & 
	 & $50\%$&$-$&$60\%$\\
$eMJ$ (RC) 	 
	 & $    147     $ & $  158.3 $ & $\pm $ & $    23.9 $   
	& $    140     $ & $  146.0 $ & $\pm $ & $    21.4 $  
	 & $10\%$&$-$&$40\%$\\
$eMJ$ (WC) 	 
	 & $      0     $ & $    1.3 $ & $\pm $ & $     0.3 $   
	& $      1     $ & $  0.6 $ & $\pm $ & $   0.4 $  
	 & $5\%$&$-$&$20\%$\\
$ee MJ$ 	 
	 & $      0     $ & $    1.5 $ & $\pm $ & $     0.5 $   
	& $      2     $ & $    1.7 $ & $\pm $ & $     0.5 $  
	 & $5\%$&$-$&$35\%$\\
$e\mu MJ$ 	 
	 & $      0     $ & $  0.03 $ & $\pm $ & $   0.02 $   
	& $      0     $ & $  0.03 $ & $\pm $ & $   0.03 $  
	 & $5\%$&$-$&$15\%$\\
$e \nu MJ$ 	 
	 & $      3     $ & $    5.6 $ & $\pm $ & $     1.2 $   
	& $      5     $ & $    8.2 $ & $\pm $ & $     2.0 $  
	 & $5\%$&$-$&$40\%$\\
$\nu MJ$ 	 
	 & $     204     $ & $   235.5 $ & $\pm $ & $     63.3 $   
	& $     113     $ & $   134.0 $ & $\pm $ & $     33.8 $  
	 & $5\%$&$-$&$15\%$\\
$\nu\mu MJ$ 	 
	 & $      0     $ & $  0.04 $ & $\pm $ & $   0.02 $    
	& $      0     $ & $  0.06 $ & $\pm $ & $   0.03 $  
	 & $5\%$&$-$&$20\%$\\

\hline

 \end{tabular}
 \end{center}
  \caption[Selection summary: total event numbers] {Total numbers of selected events, SM 
expectations and ranges of signal efficiencies for the squark decay channels considered 
in $e^-p$ and in $e^+p$ collisions. 
The range of signal efficiencies gives the
extreme values for squark
masses ranging from $100$~GeV to $290$~GeV and gaugino masses ranging from $30$~GeV up to the squark mass. 

The $\nu q$ channel is not relevant for $e^+p$ data since the $\tilde{u}_L$--type squarks produced in $e^+p$
do not undergo this decay. Only $\tilde{d}_R$--type squarks, which are produced dominantly in $e^-p$ collisions, can undergo
direct decay leading to a $\nu q$ final state.
The total error on the SM prediction is determined by adding the effects of all model and experimental systematic
uncertainties in quadrature.\\
}

  \label{tab:totnum}
\end{table}

\clearpage

\clearpage

\clearpage

\clearpage

\clearpage

\clearpage

\clearpage

\clearpage

\clearpage

\clearpage

\clearpage

\clearpage

\end{document}